%% file: blazeit-vldb.tex
\documentclass{vldb}

\newcommand{\subparagraph}{}

\usepackage{graphicx}
\usepackage{balance}  
\usepackage[dvipsnames]{xcolor}
\usepackage{xspace}
\usepackage{multirow}
\usepackage{url}
\usepackage{newfloat}
\usepackage[noabbrev]{cleveref}
\usepackage[labelfont=bf]{caption}
\usepackage[labelfont=bf]{subcaption}
\usepackage{enumitem}
\usepackage{textcomp}
\usepackage[]{algorithm2e}
\usepackage{cite}
\usepackage{upquote}
\usepackage{etoolbox}
\usepackage{listings}
\usepackage[small,compact]{titlesec}
\usepackage[utf8]{inputenc}
\usepackage{times}

\newcommand{\specialcell}[2][c]{%
    \begin{tabular}[#1]{@{}c@{}}#2\end{tabular}}

\newcommand{\minihead}[1]{{\vspace{.4em}\noindent\textbf{#1.} }}
\newcommand{\miniheadit}[1]{{\vspace{.4em}\noindent\textit{#1.} }}

\newcommand{\sntext}{BLAZEIT\xspace}
\newcommand{\sn}{\textsc{BlazeIt}\xspace}
\newcommand{\noscope}{\textsc{NoScope}\xspace}

\newcommand{\fql}{\textsc{FrameQL}\xspace}

\newtoggle{rcolors} \togglefalse{rcolors}

\lstdefinelanguage{SQL}{
 keywords={AND, OR, WHERE, SELECT, FROM, LIKE, GROUP, BY, NOT, COUNT, DISTINCT, DIFF, BETWEEN},
 keywordstyle=\color{black},
 ndkeywords={NULL, true, false},
 ndkeywordstyle=\color{BrickRed}\bfseries,
 basicstyle=\small\ttfamily,
 identifierstyle=\color{black},
 sensitive=false,
 comment=[l]{--},
 morecomment=[s]{/*}{*/},
 commentstyle=\color{NavyBlue}\ttfamily,
 string=[s]{"}{"},
 morestring=[s]{`}{'},
 showstringspaces=false,
 stringstyle=\color{violet}\ttfamily,
}
\vldbTitle{BlazeIt: Optimizing Declarative Aggregation and Limit Queries for Neural Network-Based Video Analytics}
\vldbAuthors{Daniel Kang, Peter Bailis, Matei Zaharia}
\vldbNumber{4}
\vldbVolume{13}
\vldbYear{2019}
\vldbDOI{https://doi.org/10.14778/3372716.3372725}

\begin{document}


\title{BlazeIt: Optimizing Declarative Aggregation and Limit Queries for Neural Network-Based Video Analytics}

\numberofauthors{3} 

\author{
Daniel Kang, Peter Bailis, Matei Zaharia\\
\affaddr{Stanford DAWN Project, \texttt{blazeit@cs.stanford.edu}}
}

\maketitle

\input{tex/abstract}
\input{tex/intro}

\input{tex/overview}
\input{tex/query_language}
\input{tex/operators}
\input{tex/implementation}
\input{tex/evaluation}

\input{tex/related_work}
\input{tex/conclusion}

\subsubsection*{Acknowledgements}
{
\scriptsize
This research was supported in part by affiliate members and other supporters of
the Stanford DAWN project---Ant Financial, Facebook, Google, Infosys, Intel,
NEC, SAP, Teradata, and VMware---as well as Toyota Research Institute, Keysight
Technologies, Amazon Web Services, Cisco, and the NSF under CAREER grant
CNS-1651570. Any opinions, findings, and conclusions or recommendations
expressed in this material are those of the authors and do not necessarily
reflect the views of the NSF.
}

\balance

\bibliographystyle{abbrv}
\bibliography{blazeit-vldb}

\clearpage

\input{appendix}

\end{document}

%% file: tex/abstract.tex
\begin{abstract}

\begin{sloppypar}
Recent advances in neural networks (NNs) have enabled automatic querying of
large volumes of video data with high accuracy. While these deep NNs can produce
accurate annotations of an object's position and type in video, they are
computationally expensive and require complex, imperative deployment code to
answer queries.  Prior work uses approximate filtering to reduce the cost of
video analytics, but does not handle two important classes of queries,
aggregation and limit queries; moreover, these approaches still require complex code to
deploy. To address the computational and usability challenges of querying video
at scale, we introduce \sn, a system that optimizes queries of spatiotemporal
information of objects in video. \sn accepts queries via \fql, a declarative
extension of SQL for video analytics that enables video-specific query
optimization. We introduce two new query optimization techniques in \sn that are
not supported by prior work. First, we develop methods of using NNs as control
variates to quickly answer approximate aggregation queries with error bounds.
Second, we present a novel search algorithm for cardinality-limited video
queries. Through these these optimizations, \sn can deliver up to 83$\times$
speedups over the recent literature on video processing.
\end{sloppypar}

\end{abstract}

%% file: tex/intro.tex
\section{Introduction}

\begin{sloppypar}
Two trends have caused recent interest in video analytics. First, cameras are
now cheap and widely deployed, e.g., London alone has over 500,000
CCTVs~\cite{bbc2015camera}. Second, deep neural networks (DNNs) can
automatically produce annotations of video. For example, object detection
DNNs~\cite{felzenszwalb2010object} return a set of bounding boxes and
object classes given an image or frame of video. Analysts can use these DNNs to
extract object positions and types from every frame of video, a common analysis
technique~\cite{redmon2017yolo9000}. In this work, we study the batch setting,
in which large quantities of video are collected for later
analysis~\cite{kang2017noscope, hsieh2018focus, anderson2018predicate}.
\end{sloppypar}

While DNNs are accurate~\cite{he2017mask}, naively employing them has two key
challenges. First, from a usability perspective, these methods
require complex, imperative programming across many low-level libraries, such
as OpenCV, Caffe2, and Detectron~\cite{Detectron2018}---an ad-hoc, tedious
process. Second, from a computational perspective, the naive method of
performing object detection on every frame of video is cost prohibitive at scale:
state-of-the-art object detection, e.g., Mask R-CNN~\cite{he2017mask}, runs at 3
frames per second (fps), which would take 8 GPU-decades to process 100
camera-months of video.

\begin{sloppypar}
Researchers have recently proposed optimizations for video analytics, largely focusing on
filtering via approximate predicates~\cite{kang2017noscope,
anderson2018predicate, lu2018accelerating, canel2019scaling}. For example,
\noscope and \textsc{Tahoma} train cheaper, proxy models for
filtering~\cite{kang2017noscope, anderson2018predicate}. However, these
optimizations do not handle two key classes of queries: aggregate and limit
queries. For example, an analyst may want to count the average number of cars
per frame (aggregate query) or manually inspect only 10 instances of a bus and
five cars (limit query) to understand congestion. Approximate filtering is
inefficient for these queries, e.g., filtering for cars will not significantly
speed up counting cars if 90\% of the frames contain cars. Furthermore, these
optimizations still require non-expert users to write complex code to deploy.

To address these usability and computational challenges, we present \sn, a video
analytics system with a declarative query language and two novel optimizations for
aggregation and limit queries. \sn's declarative query language, \fql, extends
SQL with video-specific functionality and allows users familiar with SQL to
issue video analytics queries. Since queries are expressed declaratively, \sn
can \emph{automatically} optimize them end-to-end with its query optimizer and
execution engine. Finally, \sn provides two novel optimizations for aggregation
and limit queries that outperforms prior work, including
\noscope~\cite{kang2017noscope} and approximate query processing (AQP), by up to
83$\times$.

\begin{figure*}[t!]
  \begin{subfigure}[h]{0.49\linewidth}
    \includegraphics[width=0.99\columnwidth]{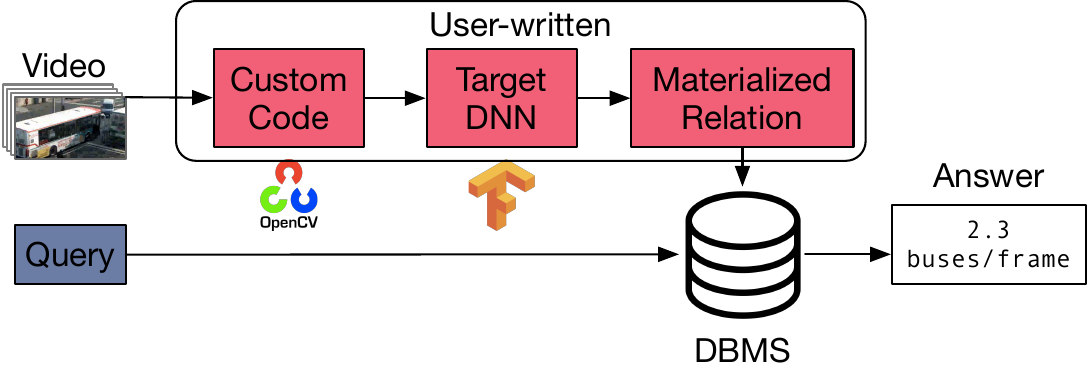}
    \caption{Schematic of the naive method of querying video. Naively using DNNs
    or human annotators is too expensive for many applications.}
    \label{fig:arch-naive}
  \end{subfigure}
  \hspace{0.01\linewidth}
  \begin{subfigure}[t]{0.49\linewidth}
    \includegraphics[width=0.99\columnwidth]{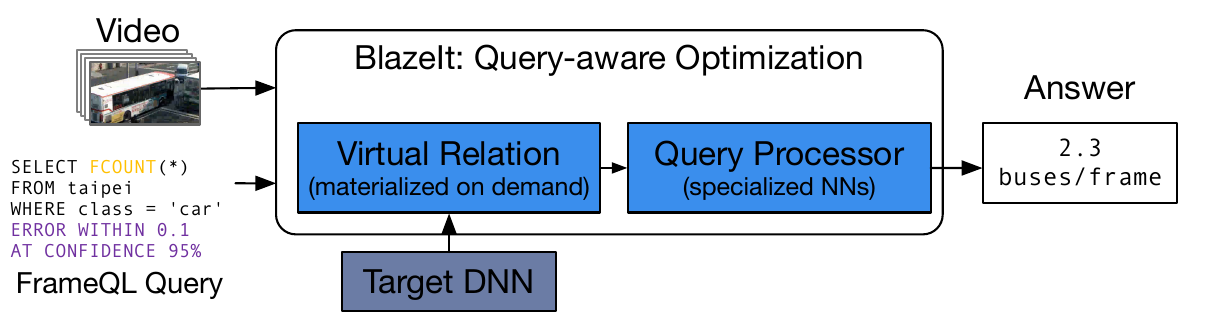}
    \vspace{3em}
    \caption{Schematic of \sn. \sn will create optimized query plans and avoid
    calling the expensive DNN where possible.}
    \label{fig:arch-blaze}
  \end{subfigure}
  \caption{Schematic of the naive method of querying video and \sn. \sn does not
  require writing complex code and does not require pre-materializing all the
  tuples.}
  \label{fig:arch-diagram}
\end{figure*}

\fql allows users to query information of objects in video through a
\emph{virtual} relation. Instead of fully materializing the \fql relation, \sn
uses optimizations to reduce the number of object detection invocations while
meeting an accuracy guarantee based on the specification of the \fql query
(Figure~\ref{fig:arch-diagram}). \fql's relation represents the information of
positions and classes of objects in the video. Given this relation, \fql can
express selection queries in prior work~\cite{kang2017noscope,
anderson2018predicate, lu2018accelerating, canel2019scaling}, along with new
classes of queries, including aggregation and limit queries (\S\ref{sec:fql}).
\end{sloppypar}

Our first optimization, to answer aggregation queries, uses query-specific NNs
(i.e., specialized NNs~\cite{kang2017noscope}) as a control variate or to
directly answer queries (\S\ref{sec:opt_agg}). Control variates are a variance
reduction technique that uses an auxiliary random variable that is correlated
with the statistic of interest to reduce the number of samples necessary for a
given error bound~\cite{hammersley1964general}. We show how to use specialized
NNs as a control variate, a novel use of specialized NNs (which have been used
for approximate filtering). In contrast, standard random sampling does not
leverage proxy models and prior work (filtering) is inefficient when objects
appear frequently.

Our second optimization, to answer cardinality-limited queries (e.g., a
\texttt{LIMIT} query searching for 10 frames with at least three cars),
evaluates object detection on frames that are more likely to contain the event
using proxy models (\S\ref{sec:op_scrubbing}). By prioritizing frames to search
over, \sn can achieve exact answers while speeding up query execution. In
contrast, filtering is inefficient for frequent objects and random sampling is
inefficient for rare events.

Importantly, both of our optimizations provide exact answers or accuracy
guarantees \emph{regardless of the accuracy of the specialized NNs}.
Furthermore, both of these optimizations can be extended to account for query
predicates.

\begin{sloppypar}
\sn incorporates these optimizations in an end-to-end system with a rule-based
query optimizer and execution engine that efficiently executes \fql queries.
Given query contents, \sn will generate an optimized query plan that avoids
executing object detection wherever possible, while maintaining the
user-specified accuracy (relative to the object detection method as ground
truth).
\end{sloppypar}

We evaluate \sn on a variety of queries on four video streams that are widely
used in studying video analytics~\cite{kang2017noscope, hsieh2018focus,
jiang2018chameleon, xu2019vstore, canel2019scaling} and two new video streams.
We show that \sn can achieve up to 14$\times$ and 83$\times$ improvement over
prior work in video analytics and AQP for aggregation and limit queries
respectively.

In summary, we make the following contributions:
\begin{enumerate}[itemsep=0em,parsep=0em,topsep=0em]
  \item We introduce \fql, a query language for spatiotemporal information of
  objects in videos, and show it can answer a variety of real-world queries,
  including aggregation, limit, and selection queries.

  \item We introduce an aggregation algorithm that uses control
  variates to leverage specialized NNs for more efficient aggregation
  than existing AQP methods by up to 14$\times$.

  \item We introduce an algorithm for limit queries that uses specialized
  NNs and can deliver up to 83$\times$ speedups over recent work in video
  analytics and random sampling.
\end{enumerate}

\begin{figure}[t!]
\centering
\begin{subfigure}[t]{0.45\columnwidth}
  \centering
  \small
  \begin{verbatim}
SELECT FCOUNT(*)
FROM taipei
WHERE class = 'car'
ERROR WITHIN 0.1
AT CONFIDENCE 95%
  \end{verbatim}
  \vspace{-0.5em}
  \caption{The \fql query for counting the frame-averaged number of cars within
  a specified error and confidence.}
  \label{query:aggregate}
\end{subfigure}
\hspace{0.02\columnwidth}
\begin{subfigure}[t]{0.49\columnwidth}
  \centering
  \small
  \begin{verbatim}
SELECT timestamp
FROM taipei
GROUP BY timestamp
HAVING SUM(class='bus')>=1
   AND SUM(class='car')>=5
LIMIT 10 GAP 300
  \end{verbatim}
  \vspace{-1.7em}
  \caption{The \fql query for selecting 10 frames of at least one bus and five
  cars, with each frame at least 300 frames apart (10s at 30 fps).}
  \label{query:scrubbing}
\end{subfigure}

\begin{subfigure}[t]{0.99\columnwidth}
  \vspace{0.5em}
  \centering
  \small
  \begin{verbatim}
SELECT *
FROM taipei
WHERE class = 'bus' AND redness(content) >= 17.5
  AND area(mask) > 100000
GROUP BY trackid HAVING COUNT(*) > 15
  \end{verbatim}
  \vspace{-1.7em}
  \caption{The \fql query for selecting all the information of red buses at
  least 100,000 pixels large, in the scene for at least 0.5s (15 frames). The
  last constraint is for noise reduction.}
  \label{query:red_bus}

\end{subfigure}
\caption{Three \fql example queries. As shown, the syntax is largely standard
SQL.}
\vspace{-0.5em}
\label{fig:fql_queries}
\end{figure}

%% file: tex/overview.tex
\section{Example Use Cases}
\label{sec:overview}

\begin{figure}[t!]
  \centering
  \begin{subfigure}{.49\columnwidth}
    \includegraphics[width=0.95\columnwidth]{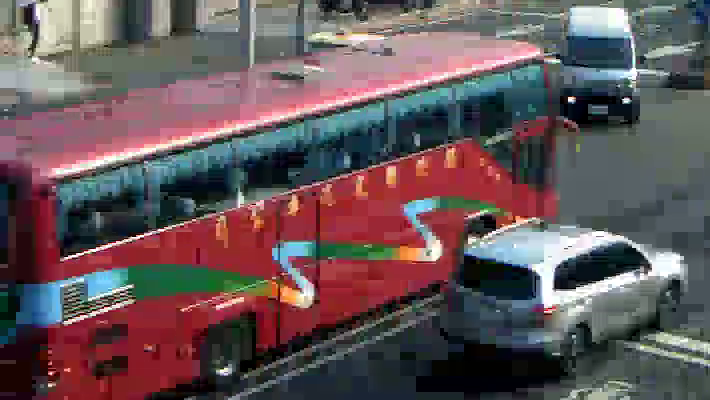}
    \vspace{-0.2em}
    \caption{Red tour bus.}
  \end{subfigure}
  \begin{subfigure}{.49\columnwidth}
    \includegraphics[width=0.95\columnwidth]{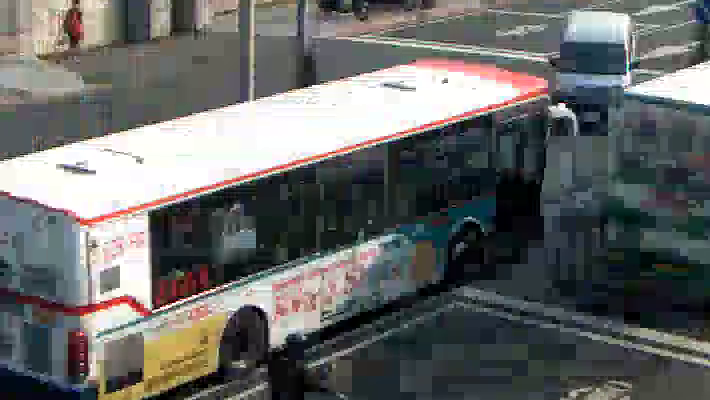}
    \vspace{-0.2em}
    \caption{White transit bus.}
  \end{subfigure}
  \vspace{-0.9em}
  \caption{Examples of buses in \texttt{taipei}. A city planner might be
  interested in distinguishing tour buses from transit buses and uses color as a
  proxy.}
  \label{fig:buses}
\end{figure}

Recall that we focus on the batch setting in this work.  We give several
scenarios where \sn could be applied:

\minihead{Urban planning}
Given a set of traffic cameras at street corners, an urban planner performs
traffic metering based on the number of cars, and determines the busiest
times~\cite{sun2003highway}. The planner is interested in how public transit
interacts with congestion~\cite{de1993transit} and looks for 10 events of at
least one bus and at least five cars. Then, the planner seeks to understand how
tourism affects traffic and looks for red buses as a proxy for tour buses (see
Figure~\ref{fig:buses}).

\minihead{Autonomous vehicle analysis}
An analyst studying AVs notices anomalous behavior when the AV
is in front of a yellow light and there are multiple pedestrians in the
crosswalk~\cite{geiger2012we}, and searches for such events.

\minihead{Store planning}
A retail store owner places a CCTV in the store~\cite{senior2007video}. The
owner segments the video into aisles and counts the number of people that walk
through each aisle to understand the flow of customers. This information can be
used for planning store and aisle layout.

\minihead{Ornithology}
An ornithologist (a scientist who studies birds) is interested in
understanding bird feeding patterns, so places a webcam in front of a bird
feeder~\cite{cornellwebcam}. Then, the ornithologist puts different bird
feed on the left and right side of the feeder and counts the number of birds
that visit each side. Finally, as a proxy for species, the ornithologist might
then select red or blue birds.

\vspace{0.2em}
These queries can be answered using spatiotemporal information of objects in the
video, along with simple user-defined functions (UDFs) over the content of the boxes. Thus, these
applications illustrate a need for a unified method of expressing such queries.

\section{\sntext System Overview}
\label{sec:overview_arch}

\sn's goal is to execute \fql queries as quickly as possible; we describe \fql
in \S\ref{sec:fql}. To execute \fql queries, \sn uses a \emph{target
object detection method}, an entity resolution method, and the optional
user-defined functions (UDFs). We describe the specification of these methods in
this section and describe our defaults in \S\ref{sec:implemenation}.
Importantly, we assume the object detection class types are provided.

\sn executes queries quickly by avoiding materialization using the techniques
described \S\ref{sec:opt_agg} and \S\ref{sec:op_scrubbing}. \sn uses proxy
models, typically specialized neural networks~\cite{kang2017noscope,
shen2017fast}, to avoid materialization (Figure~\ref{fig:arch-blaze}), which we
describe below.

%

%

\subsection{Components}

\minihead{Configuration}
We assume the target object detection method is implemented with the following
API:
\begin{equation}
  \texttt{OD(frame)} \to \texttt{Set<Tuple<class, box>>}
  \label{eq:od-def}
\end{equation}
and the object classes (i.e., types) are provided. We assume the entity
resolution takes two nearby frames and boxes and returns true if the boxes
correspond to the same object. While we provide defaults
(Section~\ref{sec:implemenation}), the object detection and entity resolution
methods can be changed, e.g., a license plate reader could be used for resolving
the identity of cars. The UDFs can be used to answer more complex queries, such
as determining color, filtering by object size or location, or fine-grained
classification. UDFs are functions that accept a timestamp, mask, and
rectangular set of pixels. For example, to compute the ``redness'' of an object,
a UDF could average the red channel of the pixels.

\minihead{Target-model annotated set (TMAS)}
At ingestion time, \sn will perform object detection over a small sample of
frames of the video with the target object detection NN and will store the
metadata as \fql tuples, which we call the \emph{target-model annotated set
(TMAS)}. This procedure is done when the data is ingested and
not per query, namely it is performed once, offline, and shared for multiple
queries later. For a given query, \sn will use this metadata to materialize
training data to train a query-specific proxy model; details are given in
\S\ref{sec:opt_agg} and \S\ref{sec:op_scrubbing}. The TMAS is split
into training data and held-out data.

\minihead{Proxy models and specialized NNs}
\sn can infer proxy models and/or filters from query predicates, many of which
must be trained from data. These proxy models can be used to accelerate query
execution \emph{with accuracy guarantees}.

Throughout, we use specialized NNs~\cite{kang2017noscope, shen2016fast},
specifically a miniaturized ResNet~\cite{he2016deep}
(\S\ref{sec:implemenation}), as proxy models. A specialized NN is a NN that
mimics a larger NN (e.g., Mask R-CNN) on a simplified task, e.g., on a marginal
distribution of the larger NN. As specialized NNs predict simpler output, they
can run dramatically faster.

\sn will infer if a specialized NN can be trained from the query specification.
For example, to replicate \noscope's binary detection, \sn would infer that
there is a predicate for whether or not there is an object of interest in the
frame and train a specialized NN to predict this. Prior work has used
specialized NNs for binary detection~\cite{kang2017noscope, han2016mcdnn}, but
we extend specialization for aggregation and limit queries.

\subsection{Limitations}
\label{sec:limitations}

While \sn can answer a significantly wider range of video queries than prior
work, we highlight several limitations.

\minihead{TMAS}
\sn requires the object detection method to be run over a portion of the data
for training specialized NNs and filters as a preprocessing step. Other
contemporary systems also require a TMAS~\cite{kang2017noscope, hsieh2018focus}.

\minihead{Model Drift}
\sn targets on the batch analytics setting where we can sample the TMAS
i.i.d.~from the available data. However, in the streaming setting, where the
data distribution may change, \sn will still provide accuracy guarantees but
performance may be reduced. Namely, the accuracy of \sn's specialized NNs may
degrade relative to the target model. As a result, \sn may execute queries more
slowly, but importantly, this will not affect accuracy
(\S\ref{sec:opt-overview}). This effect can be mitigated by labeling a portion
of new data and monitoring drift or continuous retraining.

\minihead{Object detection}
\sn depends on the target object detection method and does not
support object classes beyond what the method returns, e.g., the
pretrained Mask R-CNN~\cite{he2017mask,Detectron2018} can detect
cars, but cannot distinguish between sedans and SUVs. However, users can
supply UDFs if necessary.



%% file: tex/query_language.tex
\section{FrameQL: Expressing Complex Spatiotemporal Visual Queries}
\label{sec:fql}

\begin{table}
  \small
  \caption{\fql's data schema contains spatiotemporal and content information related
  to objects of interest, as well as metadata (class, identifiers). Each tuple
  represents an object appearing in one frame; thus a frame may have many or no
  tuples. The features can be used for downstream tasks.}
  \vspace{-0.5em}
  \label{table:fql}
  \centering
  \begin{tabular}{lll}
    Field               & Type            & Description \\ \hline
    \texttt{timestamp}  & float           & Time stamp \\
    \texttt{class}      & string          & Object class (e.g., bus, car) \\
    \texttt{mask}       & (float, float)* & Polygon containing the object \\
                        &                 & of interest, typically a rectangle \\
    \texttt{trackid}    & int             & Unique identifier for a continuous \\
                        &                 & time segment when the \\
                        &                 & object is visible \\
    \texttt{content}    & float*          & Content of pixels in mask \\
    \texttt{features}   & float*          & The feature vector output by the \\
                        &                 & object detection method.
  \end{tabular}
  \vspace{-0.5em}
\end{table}

To address the need for a unifying query language over video analytics, we
introduce \fql, an extension of SQL for querying spatiotemporal information of
objects in video. By providing a table-like schema using the standard relational
algebra, we enable users familiar with SQL to query videos, whereas
implementing these queries manually would require expertise in deep learning,
computer vision, and programming.

\fql is inspired by prior query languages for video analytics~\cite{lu2015svql,
kuo2000content, donderler2005bilvideo, le2008query}, but \fql specifically
targets information that can be populated automatically using computer vision
methods. We discuss differences in detail at the end of this section.

\minihead{\fql data model} \fql represents videos (possibly
compressed in formats such as H.264) as virtual relations, with one relation per
video. Each \fql tuple corresponds to a single object in a frame. Thus, a
frame can have zero or more tuples (i.e., zero or more objects), and the same
object can have one or more tuples associated with it (i.e., appear in several
frames).

\begin{sloppypar}
We show \fql's data schema in Table~\ref{table:fql}. It contains
fields relating to the time, location, object class, and object identifier, the box
contents, and the features from the object detection method. \sn can
automatically populate \texttt{mask}, \texttt{class}, and \texttt{features} from
the object detection method (see Eq.~\ref{eq:od-def}), \texttt{trackid} from the
entity resolution method, and \texttt{timestamp} and \texttt{content} from the
video metadata. Users can override the default object detection and entity
resolution methods. For example, an ornithologist may use an object detector
that can detect different species of birds, but an autonomous vehicle analyst
may not need to detect birds at all.\footnote{\sn will inherit any errors from
the object detection and entity resolution methods.}
\end{sloppypar}

%
%
%
%
%

\begin{figure}[t!]
\footnotesize
\begin{lstlisting}[frame=single]
SELECT * | expression [, ...]
  FROM table_name
  [ WHERE condition  ]
  [ GROUP BY expression [, ...]  ]
  [ HAVING condition [, ...]  ]
  [ LIMIT count  ]
  [ GAP count  ]
  [ ERROR WITHIN tol AT CONFIDENCE conf  ]
\end{lstlisting}
\vspace{-0.5em}
\caption{\fql syntax. As shown, \fql largely inherits SQL syntax.}
\label{fig:syntax}
\end{figure}

\minihead{\fql query format}
\fql allows selection, projection, and aggregation of objects, and, by returning
relations, can be composed with standard relational operators.  We show the \fql
syntax in Figure~\ref{fig:syntax}. \fql extends SQL in three ways: \texttt{GAP},
syntax for specifying an error tolerance (e.g., \texttt{ERROR WITHIN}), and
\texttt{FCOUNT}. Notably, we do not support joins as we do not optimize for
joins in this work, but we describe how to extend \fql with joins in an extended
version of this paper~\cite{kang2019blazeit}. We show \fql's extensions
Table~\ref{table:syntax}; several were taken from
BlinkDB~\cite{agarwal2013blinkdb}. We provide the motivation behind each
additional piece of syntax.

First, when the user selects timestamps, the \texttt{GAP} keyword ensures that
the returned frames are at least \texttt{GAP} frames apart. For example, if 10
consecutive frames contain the event and \texttt{GAP = 100}, only one frame of
the 10 frames would be returned.

Second, as in BlinkDB~\cite{agarwal2013blinkdb}, users may wish to have fast
responses to exploratory queries and may tolerate some error. Thus, we allow the
user to specify error bounds in the form of maximum absolute error, false
positive error, and false negative error, along with a specified confidence level (e.g.,
Figure~\ref{query:aggregate}). \noscope's pipeline can be replicated with \fql
using these constructs. We choose absolute error bounds in this work as the user
may inadvertanely execute a query with 0 records, which would require scanning
the entire video (\S\ref{sec:opt_agg}).

We also provide a short-hand for returning a frame-averaged count, which we
denote as \texttt{FCOUNT}. For example, consider two videos: 1) a 10,000 frame
video with one car in every frame, 2) a 10 frame video with a car only in the
first frame. \texttt{FCOUNT} would return $1$ in the first video and $0.1$ in
the second video. As videos vary in length, this allows for a normalized way of
computing errors. \texttt{FCOUNT} can easily be transformed into a
time-averaged count. Window-based analytics can be done using the
existing \texttt{GROUP BY} keyword.


\begin{table}[t!]
  \small
  \caption{Additional syntactic elements in \fql. Some of these
  were adapted from BlinkDB~\cite{agarwal2013blinkdb}.}
  \label{table:syntax}
  \centering
  \begin{tabular}{l|l}
    \specialcell{Syntactic\\element}     & Description \\ \hline
    \texttt{FCOUNT}       & Frame-averaged count (equivalent to \\
                          & time-averaged count), i.e., \\
                          & \texttt{COUNT(*) / MAX(timestamp)} \\
    \texttt{ERROR WITHIN} & Absolute error tolerance \\
    \texttt{FPR WITHIN}   & Allowed false positive rate \\
    \texttt{FNR WITHIN}   & Allowed false negative rate\\
    \texttt{CONFIDENCE}   & Confidence level \\
    \texttt{GAP}          & Minimum distance between returned frames
  \end{tabular}
  \vspace{-0.5em}
\end{table}

\minihead{\fql examples}
We describe how the some of the example use cases from
\S\ref{sec:overview} can be written in \fql. We
assume the video is recorded at 30 fps.

Figure~\ref{query:aggregate} shows how to count the average number of cars in a
frame. The query uses \texttt{FCOUNT} as the error bounds are computed
per-frame. Figure~\ref{query:scrubbing} shows how to select frames with at least
one bus and at least five cars, which uses the \texttt{GAP} keyword to
ensure events are a certain time apart. At 30 fps, \texttt{GAP 300} corresponds
to 10 seconds. Figure~\ref{query:red_bus} shows how to exhaustively select
frames with red buses. Here, \texttt{redness} and \texttt{area} are UDFs, as
described in \S\ref{sec:overview_arch}. The other example use cases can be
answered similarly.


\minihead{Comparison to prior languages}
Prior visual query engines have proposed similar schemas, \emph{but assume that
the relation is already populated}~\cite{kuo1996content,li1997moql}, i.e., that
the data has been created through external means (typically by humans). In
contrast, \fql's relation can be automatically populated by \sn. However, as we
focus on exploratory queries in this work, \fql's schema is \emph{virtual} and
rows are only populated as necessary for the query at hand, which is similar to
an unmaterialized view. This form of laziness enables a variety of optimizations
via query planning.

%% file: tex/operators.tex
\section{Query Optimizer Overview}
\label{sec:opt-overview}

\minihead{Overview}
\sn's primary challenge is executing \fql queries \emph{efficiently}: recall
that object detection is the overwhelming bottleneck
(Table~\ref{table:method-speeds}). To optimize and execute queries, \sn inspects
query contents to see if optimizations can be applied. For example, \sn cannot
optimize aggregation queries without error bounds, but can optimize aggregation
queries with a user-specified error tolerance.

\begin{table}
\centering
\caption{A comparison of object detection methods, filters, and speeds. More
accurate object detection methods are more expensive. Specialized NNs and simple
filters are orders of magnitude more efficient than object detection methods.}
\label{table:method-speeds}
\setlength\itemsep{2em}
\begin{tabular}{lll}
  Method & mAP & FPS \\ \hline \hline
  YOLOv2~\cite{redmon2017yolo9000} & 25.4 & 80 \\
  Mask R-CNN~\cite{he2017mask}     & 45.2 & 3 \\ \hline
  Specialized NN & N/A  & 35k \\
  Decoding low-resol video & N/A & 62k \\
  Color filter   & N/A  & 100k
\end{tabular}
\end{table}


\sn leverages two novel optimizations to reduce the computational cost of object
detection, targeting aggregation (\S\ref{sec:opt_agg}) and limit queries
(\S\ref{sec:op_scrubbing}). As the filters and specialized
NNs we consider are cheap compared to the object detection methods, they are
almost always worth calling: a filter that runs at 100,000 fps would need to
filter 0.003\% of the frames to be effective (Table~\ref{table:method-speeds}).
Thus, we have found a rule-based optimizer to be sufficient in optimizing \fql
queries.

Both of \sn's novel optimizations share a key property: they still provide
accuracy guarantees despite using potentially inaccurate specialized
NNs. Specifically, both optimization will only speed up query execution and will
not affect the accuracy of queries; full details are in \S\ref{sec:opt_agg} and
\S\ref{sec:op_scrubbing}.

\sn also can optimize exhaustive selection queries with predicates by
implementing optimizations in prior work, such as using \noscope's specialized
NNs as a filter~\cite{kang2017noscope, lu2018accelerating}. As this case has
been studied, we defer the discussion of \sn's query optimization for exhaustive
selection to an extended paper~\cite{kang2019blazeit}.

\sn's rule-based optimizer will inspect the query specification to decide which
optimizations to apply. First, if the query specification contains an
aggregation keyword, e.g., \texttt{FCOUNT}, \sn will apply our novel
optimization for fast aggregation. Second, if the query specification contains
the \texttt{LIMIT} keyword, \sn will apply our novel optimization for limit
queries. Finally, for all other queries, \sn will default to applying filters
similar to \noscope's~\cite{kang2017noscope}.

\minihead{Work reuse}
In addition to our novel optimizations, \sn can reuse work by storing the
specialized NN model weights and their results. The specialized NNs \sn uses are
small, e.g., $<2$ MB, compared to the size of the video.

\vspace{0.4em}
We describe the intuition, the physical operator(s), its time complexity and
correctness, and the operator selection procedure for aggregates
(\S\ref{sec:opt_agg}) and limit queries (\S\ref{sec:op_scrubbing}) below.

\input{tex/aggregation}

\section{Optimizing Limit Queries}
\label{sec:op_scrubbing}

\minihead{Overview}
In cardinality-limited queries, the user is interested in finding a limited
number of events, (e.g., 10 events of a bus and five cars, see Figure~\ref{query:scrubbing}),
typically for manual inspection. Limit queries are especially helpful for rare
events. To answer these queries, \sn could perform object detection over every
frame to search for the events. However, if the events occurs infrequently,
naive methods of random sampling or sequential scans of the video can be
prohibitively slow (e.g., at 30 fps, an event that occurs once every 30 minutes
corresponds to a rate of $1.9\times10^{-5}$).

Our key intuition is to bias the search towards regions of the video that likely
contain the event. We use specialized NNs for biased sampling, in a similar vein
to techniques from the rare-event simulation literature~\cite{juneja2006rare}.
As an example of rare-event simulation, consider the probability of flipping 80
heads out of 100 coin flips.  Using a fair coin, the probability of encountering
this event is astronomically low (rate of $5.6 \times 10^{-10}$), but using a
biased coin with $p = 0.8$ can be orders of magnitude more efficient (rate of
$1.2 \times 10^{-4}$)~\cite{juneja2006rare}.

\minihead{Physical operator and selection}
\sn currently supports limit queries searching for at least $N$ of an object
class (e.g., at least one bus and at least five cars).  In \sn, we use
specialized NNs to bias which frames to sample:
\vspace{-0.3em}
\begin{itemize}
  \setlength\itemsep{-0.0em}
  \begin{sloppypar}
  \item If there are no instances of the query in the training set, \sn will
  default to performing the object detection method over every frame and
  applying applicable filters as in prior work~\cite{kang2017noscope} 
  (random sampling is also possible).
  \end{sloppypar}
  \item If there are examples, \sn will train a specialized NN to recognize
  frames that satisfy the query.
  \item \sn rank orders the unseen data by the confidence from the specialized NN.
  \item \sn will perform object detection in the rank order until the requested
  number of events is found.
\end{itemize}
\vspace{-0.1em}

For a given query, \sn trains a specialized NN to recognize frames that satisfy
the query. The training data for the specialized NN is generated in the same way
for aggregation queries (\S\ref{sec:opt_agg}). While we could train a specialized NN as a binary classifier of the
frames that satisfy the predicate and that do not, we have found that rare
queries have extreme class imbalance. Thus, we train the specialized NN to
predict counts
instead, which alleviates the class imbalance issue; this procedure has the
additional benefit of allowing the trained specialized NN to be reused for other
queries such as aggregation. For example, suppose the user wants to find frames
with at least one bus and at least five cars. Then, \sn trains a single
specialized NN to separately count buses and cars. \sn use the sum of the
probability of the frame having at least one bus and at least five cars as its
signal. \sn takes the most confident frames until the requested number of frames
is found.

In the case of multiple object classes, \sn trains a single NN to predict each
object class separately (e.g., instead of jointly predicting ``car" and ``bus",
the specialized NN would return a separate confidence for ``car" and ``bus"), as
this results in fewer weights and typically higher performance.

After the results are sorted, the full object detector is applied until the
requested number of events is found or all the frames are searched. If the query
contains the \texttt{GAP} keyword, once an event is found, the surrounding
\texttt{GAP} frames are ignored.

\minihead{Limit queries with multiple predicates}
As with aggregation queries, a user might issue a limit query with predicates.
If there is sufficient training data in the TMAS, \sn can execute the
procedure above. If there is not sufficient training data, \sn
will train a specialized NN to search for the most selective set of predicates
that contains enough data in a similar fashion to generating an aggregation
specialized NN.

\minihead{Correctness}
\sn performs object detection on all sampled
frames, so it always returns an exact answer. All frames will be exhaustively
searched if there are fewer events than the number requested, which will also be
exact.

\minihead{Time complexity} Denote $K$ to be the number of events the user
requested, $N$ the total number of matching events, and $F$ the total number of
frames in the video. We denote, for event $i$, $f_i$ as the frame where the
event occurred. Once an event is found, the \texttt{GAP} frames around the event
can be ignored, but this is negligible in practice so we ignore it in the
analysis.

If $K > N$, then every method must consider every frame in the video, i.e.,
$F$ frames. From here on, we assume $K \leq N$.

For sequential scans, $f_K$ frames must be examined. 

For random sampling, consider the number of frames to find a single event. In
expectation, random sampling will consider $\frac{F}{N}$ frames. Under the
assumption that $K \ll N \ll F$, then random sampling will consider
approximately $\frac{K \cdot F}{N}$ frames.

While using specialized NNs to bias the search does not guarantee faster
runtime, we show in \S\ref{sec:eval} that it empirically can reduce the
number of frames considered.

%% file: tex/aggregation.tex
\section{Optimizing Aggregates}
\label{sec:opt_agg}

\minihead{Overview}
In an aggregation query, the user is interested in some statistic over the data,
such as the average number of cars per frame; see Figure~\ref{query:aggregate}
for an example. To provide exact answers, \sn must call object detection on
every frame, which is prohibitively slow. However, if the user specifies an
error tolerance, \sn accelerate query execution using two novel optimizations.

We focus on optimizing counting the number of objects in a frame.
\sn requires
training data from the TMAS (\S\ref{sec:overview}) of the desired quantity (e.g., number
of cars) to leverage specialized NNs. If there is insufficient training data,
\sn will default to random sampling. If there is sufficient training data, \sn
will first train a specialized NN to estimate the statistic: if the
specialized NN is accurate enough, \sn can return the answer directly.
Otherwise, \sn will use specialized NNs to reduce the variance of AQP via
control variates~\cite{hammersley1964general}, requiring fewer samples. We next describe these steps in detail.

\minihead{Operator Selection}
The process above is formalized in Algorithm~\ref{alg:agg}. \sn will
process the TMAS into training data for a specialized NN by materializing
labels, i.e., counts. Given these labels, \sn first determines whether there is
sufficient training data ($>1\%$ of the data has instances of the object; this
choice will only affect runtime, not accuracy) to
train a specialized NN. In cases where the training data does not contain
enough examples of interest (e.g., a video of a street intersection is unlikely
to have bears), \sn will default to standard random sampling. We use an adaptive
sampling algorithm that respects the user's error bound but can terminate early
based on the variance of the sample~\cite{mnih2008empirical}.


When there is sufficient training data, \sn will train a specialized NN and
estimate its error rate on the held-out set. If the error is smaller than the
specified error at the confidence level, it will then execute the specialized NN
on the unseen data and return the answer directly. For specialized NN
execution, \sn will subsample at
twice minimum frequency of objects appearing; the minimum frequency is estimated
from the TMAS. Sampling at this rate, i.e., the
Nyquist rate~\cite{nyquist1928certain}, will ensure that \sn will sample all
objects. As specialized NNs are significantly faster than object detection, this
procedure results in much faster execution.

When the specialized NN is not accurate enough, it is used as a control variate:
a cheap-to-compute auxiliary variable correlated with the true
statistic. Control variates can approximate the statistic with fewer samples
than naive random sampling.

\begin{algorithm}[t]
  \KwData{TMAS, unseen video,

  $uerr\gets$ user's requested error rate,

  $conf\gets$ user's confidence level}
  \KwResult{Estimate of requested quantity}
  \eIf{training data has instances of object}{
    train specialized NN on TMAS\;
    $err\gets$ specialized NN error rate\;
    $\tau\gets$ average of specialized NN over unseen video\;
    \eIf{$P(err < uerr) < conf$}{
      return $\tau$\;
    }{
      $\hat m \gets$ result of Equation~\ref{eq:control} (control variates)\;
      return $\hat m$\;
    }
  }{
    Return result of random sampling.\;
  }
  \caption{\sn's aggregation query procedure. \sn will use specialized NNs for
  accelerated query execution via control variates or query rewriting where
  possible.}
  \label{alg:agg}
\end{algorithm}

\minihead{Physical Operators}
We describe the procedures for sampling, query rewriting, and control variates
below.

\miniheadit{Sampling}
When the query contains a tolerated error rate and there is not sufficient
training data for a specialized NN, \sn samples from the video, populating at
most a small number of rows for faster execution. Similar to online
aggregation~\cite{hellerstein1997online}, we provide absolute error bounds, but
the algorithm could be easily modified to give relative error bounds.
\sn uses Empirical Bernstein stopping (EBS)~\cite{mnih2008empirical},
which allows for early termination based on the variance, which is useful for
control variates. We specifically use Algorithm 3 in~\cite{mnih2008empirical};
we give an overview of this algorithm in an extended version of this
paper~\cite{kang2019blazeit}.

EBS provides an always valid, near-optimal stopping rule for bounded
random variables. EBS is always-valid in the sense that when EBS terminates, it
will respect the user's error bound and confidence; the guarantees come from a
union bound~\cite{mnih2008empirical}. EBS is near-optimal in the
following sense. Denote the user-defined error and confidence as
$\epsilon$ and $\delta$. Denote the range of the random variable to be $R$. EBS
will stop within $c \cdot \log \log \frac{1}{\epsilon \cdot | \mu | }$ of any
optimal stopping rule that satisfies $\epsilon$ and $\delta$. Here, $c$ is a
constant and $|\mu|$ is the mean of the random variable.

\miniheadit{Query Rewriting via Specialized NNs}
In cases where the specialized NN is accurate enough (as determined by the
bootstrap on the held-out set; the accuracy of the specialized NN depends on the
noisiness of the video and object detection method), \sn can return the answer
directly from the specialized NN run over all the frames for dramatically faster
execution and bypass the object detection entirely. \sn uses multi-class
classification for specialized NNs to count the number of objects in a frame.

To train the specialized NN, \sn selects the number of classes equal to the
highest count that is at least 1\% of the video plus one. For example, if 1\% of
the video contains 3 cars, \sn will train a specialized NN with 4 classes,
corresponding to 0, 1, 2, and 3 cars in a frame. \sn uses 150,000 frames for
training and uses a standard training procedure for NNs (SGD with
momentum~\cite{he2016deep}) for one epoch with a fixed learning rate of
0.1.

\sn estimates the error of the specialized NN on a held-out set using the
bootstrap~\cite{efron1994introduction}. If the error is low enough at the given
confidence level, \sn will process the unseen data using the specialized NN and
return the result.

\miniheadit{Control Variates}
In cases where the user has a stringent error tolerance, specialized NNs may not
be accurate enough to answer a query on their own. To reduce the cost of
sampling from the object detector, \sn introduces a novel method of using
specialized NNs while still guaranteeing accuracy. In particular, we adapt the
method of control variates~\cite{hammersley1964general} to video analytics (to
our knowledge, control variates have not been applied to database query
optimization or video analytics). Specifically, control variates is a method of
variance reduction~\cite{robert2004monte, johnson2013accelerating}) which uses a
proxy variable correlated with the statistic of interest. Intuitively, by
reducing the variance of sampling, we can reduce the number of frames that have
to be sampled and processed by the full object detector.

\begin{figure}[t]
  \centering
  \includegraphics[width=0.85\columnwidth]{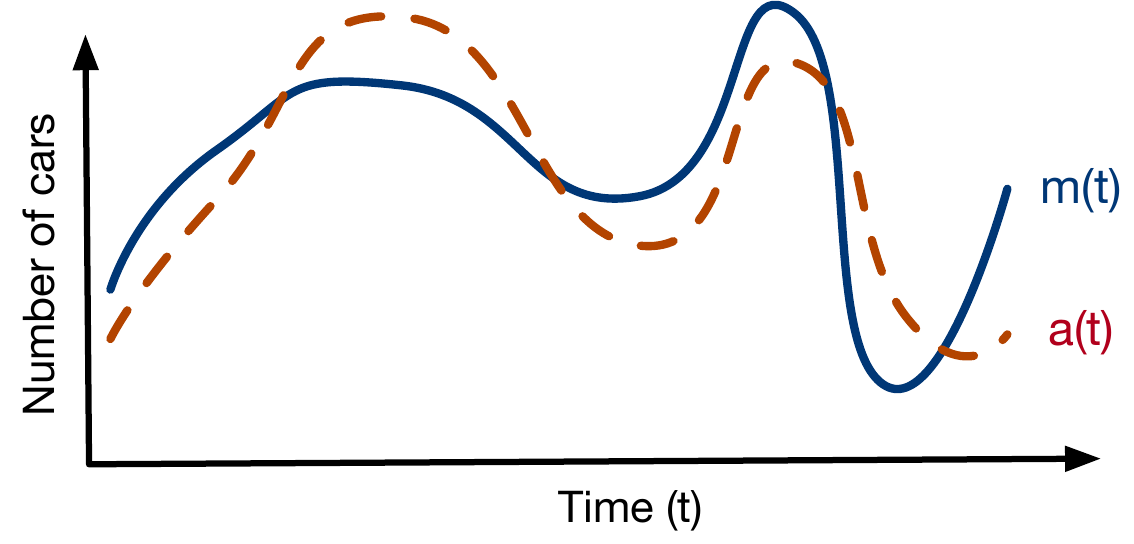}
  \caption{Schematic of control variates. Here, $a(t)$ is the result of the
  specialized NN and $m(t)$ is the result of object detection. $a$ is cheap to
  compute, but possibly inaccurate. Nonetheless, $\hat{m} = m + c \cdot (a -
  \mathbb{E}[a])$ has lower variance than $m$; thus we can use $a$ to compute
  $\mathbb{E}[m]$ with fewer samples from $m$.}
  \label{fig:cv-diagram}
\end{figure}

To formalize this intuition, suppose we wish to estimate the expectation
$m$ and we have access to an auxiliary variable $a$. The desiderata for
$a$ are that: 1) $a$ is cheaply computable, 2) $a$ is correlated with $m$ (see
time complexity). We further assume we can compute $\mathbb{E}[a] = \alpha$ and
$\mathrm{Var}(a)$ exactly. Then,
\begin{equation}
\hat m = m + c \cdot (a - \alpha)
\label{eq:control}
\end{equation}
is an unbiased estimator of $m$ for any choice of
$c$~\cite{hammersley1964general}. The optimal choice of $c$ is $c = -
\frac{\mathrm{Cov}(m, a)}{\mathrm{Var}(a)}$ and using this choice of $c$ gives
$\mathrm{Var}(\hat m) = (1 - \mathrm{Corr}(m, a)^2) \mathrm{Var}(m).$
As an example, suppose $a = m$. Then, $\hat m = m + c(m -
\mathbb{E}[m]) = \mathbb{E}[m]$ and $\mathrm{Var}(\hat m) = 0$.


This formulation works for arbitrary $a$, but choices where $a$ is correlated
with $m$ give the best results. As we show in \S\ref{sec:eval_aggregate},
specialized NNs can provide a correlated signal to the ground-truth object
detection method for all queries we consider.

As an example, suppose we wish to count the number of cars per frame; we show a
schematic in Figure~\ref{fig:cv-diagram}. Then, $m$
is the random variable denoting the number of cars the object detection method
returns. In \sn, we train a specialized NN to count the number of cars per
frame. Ideally, the specialized NN would exactly match the object detection
counts, but this is typically not the case. However, the specialized NNs are
typically correlated with the true counts. Thus, the random variable $a$ would
be the output of the specialized NN. As our choice of specialized NNs are
extremely cheap to compute, we can calculate their mean and variance exactly on
all the frames. \sn estimates $\mathrm{Cov}(m, a)$ at every round.

\minihead{Aggregation with query predicates}
A user might issue an aggregation query that contains predicates such as
filtering for large red buses (see Figure~\ref{fig:buses}). In this case, \sn
will execute a similar procedure above, but first applying the predicates to the
training data. The key difference is that in cases where there is not enough
training data, \sn will instead generate a specialized NN to count the most
selective set of predicates that contains enough data.

For example, consider a query that counts the number of large red buses. If
there is not enough data to train a specialized NN that counts the number of
large red buses, \sn will instead train a specialized NN that counts the number
of large buses (or red buses, depending on the training data). If there is no
training data for the quantity of interest, \sn will default to standard
sampling.

As control variates only requires that the proxy variable, i.e., the specialized
NN in this case, be \emph{correlated} with the statistic of interest, \sn will
return a correct answer even if it trains a specialized NN that does not
directly predict the statistic of interest.

\minihead{Correctness}
The work in \cite{mnih2008empirical} proves that EBS is an always valid,
near-optimal stopping rule. Briefly, EBS maintains an upper and lower bound of the estimate
that always respects the confidence interval and terminates when the error bound
is met given the range of the data. We estimate the range from the TMAS, which
we empirically show does not affect the confidence intervals in
Appendix~\ref{appendix:experiments}. Furthermore, while video is temporally
correlated, we assume all the video is present, namely the batch setting. As a
result, shuffling the data will result in
i.i.d.~samples. Control variates are an unbiased estimator for the statistic of
interest~\cite{hammersley1964general}, so standard proofs of correctness apply
to control variates.

Query rewriting using specialized NNs will respect the requested error bound and
confidence level under the assumption of no model drift (see
\S\ref{sec:limitations}).

\minihead{Time and sample complexity}
\sn must take $c_\delta \frac{\sigma^2}{\epsilon^2}$ samples from a random
variable with standard deviation $\sigma$ ($c_\delta$ is a constant that depends
on the confidence level and the given video). Denote the standard deviation of
random sampling as $\sigma_a$ and from control variates as $\sigma_c$; the amortized cost
of running a specialized NN on a single frame as $k_s$ and of the object
detection method as $k_o$; the total number of frames as $F$.

Control variates are beneficial when $k_s F < k_o \frac{c_\delta}{\epsilon^2}
(\sigma_a^2 - \sigma_c^2)$. Thus, as the error bound decreases or the difference
in variances increases (which typically happens when specialized NNs are more
accurate or when $\sigma_a$ is large), control variates give larger speedups.

While $\sigma_a$ and $\sigma_c$ depend on the query, we empirically show in
\S\ref{sec:eval}
that control variates and query rewriting are beneficial.

%% file: tex/implementation.tex
\section{Implementation}
\label{sec:implemenation}
We implemented our prototype of \sn in Python 3.5 for the control plane (the
deep learning frameworks we use for object detection require Python) and, 
for efficiency purposes, we implement the non-NN filters in C++. We
use PyTorch v1.0 for the training and evaluation of specialized NNs. For
object detection, we use FGFA~\cite{zhu2017flow} using MXNet v1.2 and Mask
R-CNN~\cite{he2017mask} using the Detectron framework~\cite{Detectron2018} in
Caffe v0.8. We modify the implementations to accept arbitrary parts of video.
For FGFA, we use the provided pre-trained weights and for Mask R-CNN, we use the
pretrained \texttt{X-152-32x8d-FPN-IN5k} weights. We ingest video via OpenCV.


\begin{sloppypar}
\sn uses a Fluent DSL written in Python to specify \fql queries. The cost of
storing and materializing the processed data is negligible, so we use Pandas
for processing tuples.
\end{sloppypar}

\minihead{Video ingestion}
\sn loads the video and resizes the frames to the appropriate size for each
NN (65$\times$65 for specialized NNs, short side of 600 pixels for object
detection methods), and normalizes the pixel values appropriately.

\minihead{Specialized NN training}
We train the specialized NNs using PyTorch v1.0. Video are ingested and resized
to 65$\times$65 pixels and normalized using standard ImageNet
normalization~\cite{he2016deep}. Standard cross-entropy loss is used for
training, with a batch size of 16. We use SGD with a momentum of 0.9. Our
specialized NNs use a ``tiny ResNet" architecture, a modified version of the
standard ResNet architecture~\cite{he2016deep}, which has 10 layers and a
starting filter size of 16, for all query types. As this work focuses on
exploratory queries, we choose tiny ResNet as a good default and show that it
performs better than or on par with the NNs used in~\cite{kang2017noscope}.

\minihead{Identifying objects across frames}
Our default for computing \texttt{trackid} uses motion
IOU~\cite{zhu2017flow}. Given the set of objects in two
consecutive frames, we compute the pairwise IOU of each object in the two
frames. We use a cutoff of 0.7 to call an object the same across consecutive
frames.

%% file: tex/evaluation.tex
\section{Evaluation}
\label{sec:eval}

We evaluated \sn on a variety of \fql queries on real-world video streams on
aggregate and limit queries. We show that:
\begin{enumerate}[itemsep=.1em,parsep=.1em,topsep=.1em]
  \item \sn achieves up to a
  14$\times$ speedup over AQP on aggregation queries
  (\S\ref{sec:eval_aggregate}).
  \item \sn achieves up to an 83$\times$ speedup compared to the next best
  method for video limit queries (\S\ref{sec:eval_scrubbing}).

\end{enumerate}

\subsection{Experimental Setup}

\begin{table*}[t!]
  \small
\centering
\caption{Video streams and object labels queried in our evaluation. We show the data from the test
set, as the data from the test set will influence the runtime of the baselines
and \sn.}
\vspace{-0.5em}
\label{table:videos}
\setlength\itemsep{2em}
\begin{tabular}{lllllllllll}
  \specialcell{Video name} & \specialcell{Object} &
    \specialcell{Occupancy} & \specialcell{Avg. duration\\of object in scene} &
    \specialcell{Distinct\\count} &
    \specialcell{Resol.} & \specialcell{FPS} & \specialcell{\# Eval\\frames} &
    \specialcell{Length (hrs)} & \specialcell{Detection\\method} &
    \specialcell{Thresh} \\ \hline
  \texttt{taipei}       & bus  & 11.9\% & 2.82s & 1749  & 720p  & 30 & 1188k & 33 & FGFA & 0.2\\
                        & car  & 64.4\% & 1.43s & 32367 &       &    &       &    & \\
  \texttt{night-street} & car  & 28.1\% & 3.94s & 3191  & 720p  & 30 & 973k  & 27 & Mask & 0.8\\
  \texttt{rialto}       & boat & 89.9\% & 10.7s & 5969  & 720p  & 30 & 866k  & 24 & Mask & 0.8\\
  \texttt{grand-canal}  & boat & 57.7\% & 9.50s & 1849  & 1080p & 60 & 1300k & 18 & Mask & 0.8\\
  \texttt{amsterdam}    & car  & 44.7\% & 7.88s & 3096  & 720p  & 30 & 1188k & 33 & Mask & 0.8\\
  \texttt{archie}       & car  & 51.8\% & 0.30s & 90088 & 2160p & 30 & 1188k & 33 & Mask & 0.8
\end{tabular}
\vspace{-0.8em}
\end{table*}

\begin{sloppypar}
\minihead{Evaluation queries and videos}
We evaluated \sn on six videos shown in Table~\ref{table:videos}, which were
scraped from YouTube. \texttt{taipei}, \texttt{night-street},
\texttt{amsterdam}, and \texttt{archie} are widely used in video analytics
systems~\cite{kang2017noscope, hsieh2018focus, jiang2018chameleon, xu2019vstore, canel2019scaling}
and we collected two other streams. We only considered times where the
object detection method can perform well (due to lighting conditions), which
resulted in 6-11 hours of video per day. These datasets vary in object class
(car, bus, boat), occupancy (12\% to 90\%), and average duration of object
appearances (1.4s to 10.7s). For each webcam, we use three days of video: one
day for training labels, one day for threshold computation, and one day for
testing, as in~\cite{kang2017noscope}.
\end{sloppypar}

We evaluate on queries similar to Figure~\ref{fig:fql_queries}, in which the
class and video were changed.

\minihead{Target object detection methods}
For each video, we used a pretrained object detection method as the
target object detection method, as
pretrained NNs do not require collecting additional data or training: collecting
data and training is difficult for non-experts. We selected between Mask
R-CNN~\cite{he2017mask} pretrained on MS-COCO~\cite{lin2014microsoft},
FGFA~\cite{zhu2017flow} pretrained on
ImageNet-Vid~\cite{russakovsky2015imagenet}, and
YOLOv2~\cite{redmon2017yolo9000} pretrained on MS-COCO.

\begin{sloppypar}
We labeled part of each video using Mask R-CNN~\cite{he2017mask},
FGFA~\cite{zhu2017flow}, and YOLOv2~\cite{redmon2017yolo9000}, and manually
selected the most accurate method for each video. Mask R-CNN and
FGFA are significantly more accurate than YOLOv2, so we did not select YOLOv2
for any video. The chosen object detection method per video was used for
all queries for that video.
\end{sloppypar}

In timing the naive baseline, we only included the GPU compute time and
exclude the time to process the video and convert tuples to \fql format,
as object detection is the overwhelming computational cost.


\minihead{Data preprocessing}
The literature reports that state-of-the-art object detection methods still
suffer in performance for small objects~\cite{he2017mask, zhu2017flow}. Thus, we
only considered regions where objects are large relative to the size of the frame
(these regions are video dependent). Object detectors will return a set of boxes
and confidences values. We manually selected confidence thresholds for each video
and object class for when to consider an object present
(Table~\ref{table:videos}).


\minihead{Evaluation metrics}
We computed all accuracy metrics with respect to the object detection method,
i.e., we treated the object detection method as ground truth. For aggregation
queries, we report the absolute error. For limit queries, we
guarantee only true positives are returned, thus we only report throughput.

We have found that modern object detection methods can be
accurate at the frame level. Thus, we considered accuracy at the
\emph{frame level}, in contrast to to the one-second binning that is used
in~\cite{kang2017noscope} to mitigate label flickering for \noscope.

We measured throughput by timing the complete end-to-end system excluding the
time taken to decode video, as is standard~\cite{kang2017noscope,
lu2018accelerating}. We assume the TMAS is computed offline once, so we
excluded the time to generate the TMAS. 
Unlike in~\cite{kang2017noscope}, we
also show runtime numbers \emph{when the training time of the specialized NN
is included}. We include this time as \sn
focuses on exploratory queries, whereas \noscope focuses on long-running streams
of data. We additionally show numbers where the training time is excluded, which
could be achieved if the specialized NNs were indexed ahead of time.

\minihead{Hardware Environment}
We performed our experiments on a server with an NVIDIA Tesla P100 GPU and two
Intel Xeon E5-2690v4 CPUs (56 threads). The system has 504 GB of RAM.

\subsubsection{Binary Oracle Configuration}
Many prior visual analytics systems answer binary classification
queries, including \noscope, \textsc{Tahoma}, and probablistic predicates~\cite{kang2017noscope, lu2018accelerating, hsieh2018focus} which are the
closest systems to \sn. These systems cannot directly answer queries in the form
of aggregate or limit queries for multiple instances of an object or objects.

As binary classification is not directly applicable to the tasks we consider, where relevant,
we compared against a \emph{binary oracle}, namely a method that returns (on a
frame-by-frame basis) whether or not an object class is present in the scene.
We assume the oracle is free to query. Thus, this oracle is strictly more
powerful---both in terms of accuracy and speed---than existing systems. We describe how
the binary oracle can be used to answer each type of query.

\minihead{Aggregates}
Binary oracles cannot distinguish between one and several objects, so object detection
must be performed on every frame with an object to identify the individual
objects. Thus, counting cars in \texttt{taipei} would require performing object
detection on 64.4\% of the frames, i.e., the occupancy rate.

\minihead{Cardinality-limited queries}
As above, a binary oracle can be used to filter frames that do not contain the objects
of interest. For example, if the query were searching for at least one bus and
at least five cars in \texttt{taipei}, a binary oracle can be used to remove frames
that do not have a bus and a car. Object detection will then be performed on the
remaining frames until the requested number of events is found.


\input{tex/eval_aggregate.tex}

\input{tex/eval_scrubbing.tex}

\input{tex/eval_spec.tex}

%% file: tex/eval_aggregate.tex
\subsection{Aggregate Queries}
\label{sec:eval_aggregate}

We evaluated \sn on six aggregate queries across six videos. The queries are
similar to the query in Figure~\ref{query:aggregate},
with the video and object class changed. We ran five variants of each query:
\begin{itemize}[itemsep=0em, topsep=0em]
  \item Naive: we performed object detection on every frame.
  \item Binary oracle: we performed object detection on every frame with the
  object class present.
  \item Naive AQP: we randomly sampled from the video.
  \item \sn: we used specialized NNs and control variates for efficient sampling.
  \item \sn (no train): we excluded the training time from \sn.
\end{itemize}

There are two qualitatively different execution modes: 1) where \sn rewrites the
query using a specialized NN and 2) where \sn samples using specialized NNs as
control variates (\S\ref{sec:opt_agg}). We analyzed these cases separately.

\begin{figure}[t!]
  \centering
  \includegraphics[width=0.99\columnwidth]{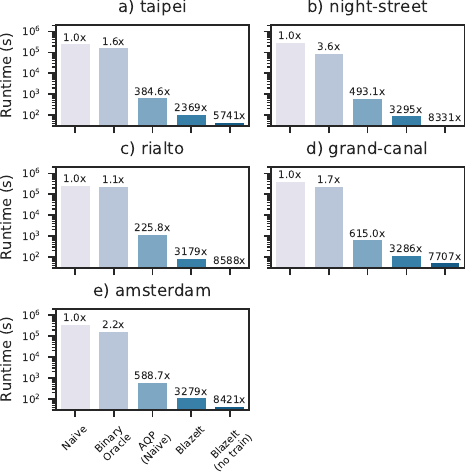}
  \vspace{-0.5em}
  \caption{
  End-to-end runtime of aggregate queries where \sn rewrote the query
  with a specialized network, measured in seconds (log scale). \sn outperforms
  all baselines. All queries targeted $\epsilon=0.1$.}
  \label{fig:agg-bypass}
\end{figure}

\minihead{Query rewriting via specialized NNs}
We evaluated the runtime and accuracy of specialized NNs when the query can be
rewritten by using a specialized NN. We ran each query with a target error rate
of 0.1 and a confidence level of 95\%. We show the average of three runs. Query
rewriting was unable to achieve this accuracy for \texttt{archie}, so we
excluded it. However, we show below that specialized NNs can be used as a
control variate even in this case.

As shown in Figure~\ref{fig:agg-bypass}, \sn can outperform naive AQP by up to
14$\times$ even when including the training time and time to compute thresholds,
which the binary oracle does not include. The binary oracle baseline does not
perform well when the video has many objects of interest (e.g.,
\texttt{rialto}).

While specialized NNs do not provide error guarantees, we show that the
absolute error stays within the 0.1 for the given videos in
Table~\ref{table:agg-bypass-err}. This shows that
specialized NNs can be used for query rewriting while respecting the user's
error bounds.

\begin{table}[t!]
  \small
  \centering
  \caption{Average error of 3 runs of query-rewriting using a specialized NN
  for counting. These videos stayed within $\epsilon = 0.1$.}
  \label{table:agg-bypass-err}
  \setlength\itemsep{2em}
  \begin{tabular}{lc}
  Video Name            & Error  \\ \hline
  \texttt{taipei}       & 0.043  \\
  \texttt{night-street} & 0.022  \\
  \texttt{rialto}       & -0.031 \\
  \texttt{grand-canal}  & 0.081 \\
  \texttt{amsterdam}    & 0.050 \\
  \end{tabular}
\end{table}

\begin{table}[t!]
  \small
  \centering
  \caption{Estimated and true counts for specialized NNs run on two different
  days of video.}
  \label{table:agg-bypass-swap}
  \setlength\itemsep{1em}
  \begin{tabular}{l|cc|cc}
  \specialcell{Video\\Name} &
    \specialcell{Pred\\(day 1)} & \specialcell{Actual\\(day 1)} &
    \specialcell{Pred\\(day 2)} & \specialcell{Actual\\(day 2)} \\ \hline
  \texttt{taipei}       & 0.86 & 0.85 & 1.21 & 1.17 \\
  \texttt{night-street} & 0.76 & 0.84 & 0.40 & 0.38 \\
  \texttt{rialto}       & 2.25 & 2.15 & 2.34 & 2.37 \\
  \texttt{grand-canal}  & 0.95 & 0.99 & 0.87 & 0.81
  \end{tabular}
  \vspace{-0.5em}
\end{table}

\minihead{Sampling and control variates}
We evaluated the runtime and accuracy of sampling with specialized NNs as a
control variate. Because of the high computational cost of running object
detection, we ran the object detection method once and recorded the results. The
run times in this section are estimated from the number of object detection
invocations.

\begin{figure}[t!]
  \centering
  \includegraphics[width=0.99\columnwidth]{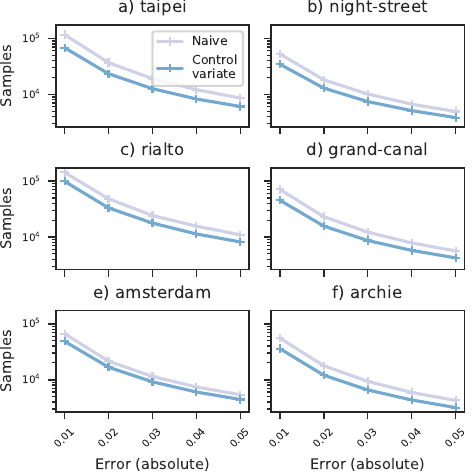}
  \vspace{-0.5em}
  \caption{
  Sample complexity of random sampling and \sn with control variates. Control
  variates via specialized NNs consistently outperforms standard random
  sampling. Note the y-axis is on a log scale.}
  \label{fig:agg-control}
\end{figure}

We targeted error rates of 0.01, 0.02, 0.03, 0.04, and 0.05 with a
confidence level of 95\%. We averaged the number of samples for each error level
over 100 runs.

As shown in Figure~\ref{fig:agg-control}, using specialized NNs as a control
variate can deliver up to a 1.7$\times$ reduction in sample complexity. As
predicted by theory, the reduction in variance depends on the correlation
between the specialized NNs and the object detection methods. Specifically, as
the correlation coefficient increases, the sample complexity decreases.

\minihead{Sampling with predicates}
We evaluated the runtime of \sn on aggregation queries with predicates. We
evaluated on one query per video and counted the number of objects with a given
color and at least a given size; full query details are give in an extended
version of this paper~\cite{kang2019blazeit}. We targeted an error rate of 0.001.

\begin{figure}[t!]
  \centering
  \includegraphics[width=0.99\columnwidth]{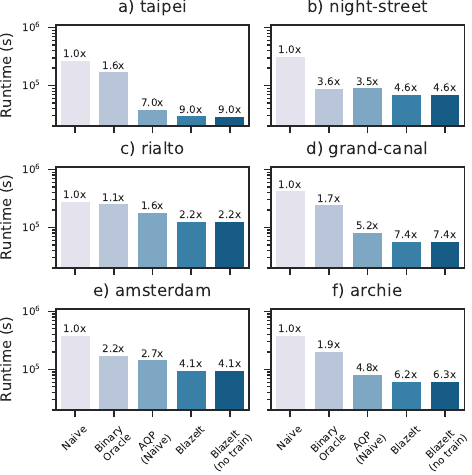}
  \vspace{-0.5em}
  \caption{Runtime of \sn and baselines for aggregation queries with predicates.
  Note the y-axis is on a log scale. As shown, \sn consistently outperforms
  naive random sampling.}
  \label{fig:agg-color}
\end{figure}

As shown in Figure~\ref{fig:agg-color}, using specialized NNs as control
variates can deliver up to a 1.5$\times$ speedup compared to naive AQP. While
the absolute runtimes vary depending on the difficulty of the query, the
relative gain of \sn's control variates only depends on the reduction in
variance. Finally, we note the gains are lower compared to queries with
predicates as there is less training data.

\minihead{Specialized NNs do not learn the average}
Specialized NNs may perform well by simply learning the average number of cars.
To demonstrate that they do not, we swapped the day of video for choosing
thresholds and testing data. We show the true counts for each day and the
average of 3 runs in Table~\ref{table:agg-bypass-swap}. Notably, we see that the
specialized NNs return different results for each day. This shows that the
specialized NNs do not learn the average and return meaningful results.

%

%% file: tex/eval_scrubbing.tex
\subsection{Cardinality-limited Queries}
\label{sec:eval_scrubbing}
We evaluated \sn on limit queries, in which frames of interest are
returned to the user, up to the requested number of frames. The queries are
similar to the query in Figure~\ref{query:scrubbing}. We show in
Table~\ref{table:scrubbing-instances} the query details and the number of
instances of each query. If the user queries more than the maximum number of
instances, \sn must inspect every frame. Thus, we chose queries with at least 10
instances of the query.

\sn will only return true positives for limit queries
(\S\ref{sec:op_scrubbing}), thus we only report the runtime. Additionally, if
we suppose that the videos are indexed with the output of the specialized NNs,
we can simply query the frames using information from the index. This scenario
might occur when the user executed an aggregate query as above. Thus, we
additionally report sample complexity.

We ran the following variants:
\begin{itemize}[itemsep=0em, topsep=0em]
  \item Naive: we performed object detection sequentially until the requested
  number of frames is found.
  \item Binary oracle: we performed object detection over the frames containing the
  object class(es) of interest until the requested number of frames is found.
  \item Sampling: we randomly sampled the video until the requested number of
  events is found.
  \item \sn: we use specialized NNs as a proxy signal to rank the frames
  (\S\ref{sec:op_scrubbing}).
  \item \sn (indexed): we assume the specialized NN has been trained and run
  over the remaining data, as might happen if a user runs several queries about
  some class.
\end{itemize}

\begin{table}[t!]
  \small
  \centering
  \caption{Query details and number of instances. We selected rare events with
  at least 10 instances.}
  \vspace{-0.5em}
  \label{table:scrubbing-instances}
  \setlength\itemsep{2em}
  \begin{tabular}{lccc}
  Video name            & Object & Number & Instances \\ \hline
  \texttt{taipei}       & car    & 6      & 70 \\
  \texttt{night-street} & car    & 5      & 29 \\
  \texttt{rialto}       & boat   & 7      & 51 \\
  \texttt{grand-canal}  & boat   & 5      & 23 \\
  \texttt{amsterdam}    & car    & 4      & 86 \\
  \texttt{archie}       & car    & 4      & 102
  \end{tabular}
  \vspace{-0.5em}
\end{table}

\begin{figure}[t!]
  \centering
  \includegraphics[width=0.99\columnwidth]{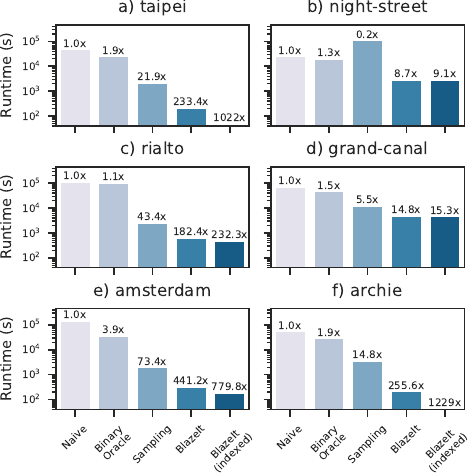}
  \vspace{-0.5em}
  \caption{
  End-to-end runtime of baselines and \sn on limit queries; \sn outperforms all
  baselines. The y-axis is log-scaled. All queries looked for 10 events.}
  \label{fig:scrubbing-all}
\end{figure}

\minihead{Single object class}
Figure~\ref{fig:scrubbing-all} shows that \sn can achieve over a 1000$\times$
speedup compared to baselines. We see that the baselines do poorly in finding
rare objects, where \sn's specialized NNs can serve as a high-fidelity signal.

\begin{figure}[t!]
  \centering
  \includegraphics[width=0.99\columnwidth]{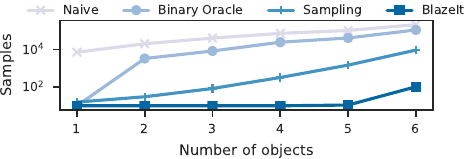}
  \vspace{-0.5em}
  \caption{
  Sample complexity of baselines and \sn when searching for at least $N$ cars in
  \texttt{taipei}; \sn outperforms all baselines. Note the y-axis is on a
  log-scale. All queries searched for 10 events.
  }
  \label{fig:scrubbing-nb-objs}
\end{figure}

We also varied the number of cars in \texttt{taipei} to see if \sn could also
search for common objects. As shown in Figure~\ref{fig:scrubbing-nb-objs}, the
sample complexity increases as the number of cars increases for both the naive
method and the binary oracle. However, for up to 5 cars, \sn's sample complexity remains
nearly constant, which demonstrates the efficacy of biased sampling. While \sn
shows degraded performance with 6 cars, there are only 70 such instances, and is
thus significantly harder to find.

\minihead{Multiple object classes}
We tested \sn on multiple object classes by searching for at least one bus and at
least five cars in \texttt{taipei}. There are 63 instances in the test set. 

\begin{figure}[t!]
  \centering
  \includegraphics[width=0.85\columnwidth]{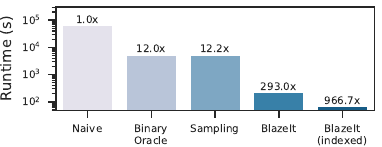}
  \vspace{-0.5em}
  \caption{
  End-to-end runtime of baselines and \sn on finding at least one bus and at
  least five cars in \texttt{taipei}; \sn outperforms all baselines. Note the
  y-axis is on a log scale.
  }
  \label{fig:scrubbing-multi}
\end{figure}

As shown in Figure~\ref{fig:scrubbing-multi}, \sn outperforms the naive baseline
by up to 966$\times$. Searching for multiple object classes is favorable for the
binary oracle, as it becomes more selective. Nonetheless, \sn significantly
outperforms the binary oracle, giving up to a 81$\times$ performance increase.

\begin{figure}[t!]
  \centering
  \includegraphics[width=0.99\columnwidth]{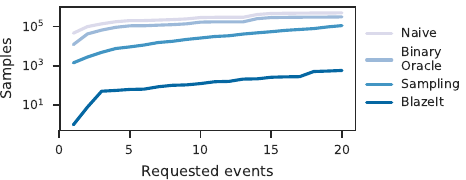}
  \vspace{-0.5em}
  \caption{
  Sample complexity of \sn and baselines when searching for at least one bus and
  at least five cars in \texttt{taipei}; \sn outperforms all baselines. The
  x-axis is the number of requested frames. Note the y-axis is on a log scale.
  }
  \label{fig:scrubbing-taipei}
\end{figure}

Additionally, we show the sample complexity as a function of the \texttt{LIMIT}
in Figure~\ref{fig:scrubbing-taipei} of \sn and the baselines, for
\texttt{taipei}. We see that \sn can be up to orders of magnitude more sample
efficient over both the naive baseline and the binary oracle.

\minihead{Limit queries with predicates}
We evaluated \sn on limit queries with predicates by searching for objects with
a specified color and at least a specified size. We present the full query details
and statistics in an extended version of this paper~\cite{kang2019blazeit}.

\begin{figure}[t!]
  \centering
  \includegraphics[width=0.99\columnwidth]{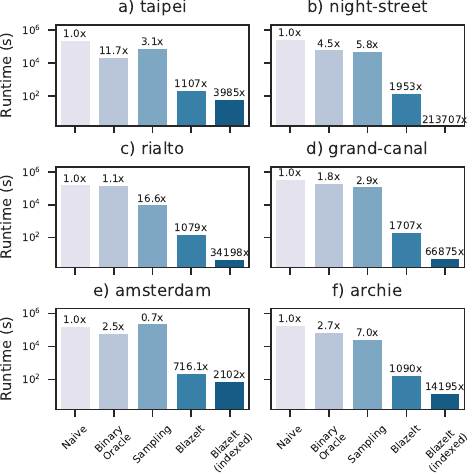}
  \vspace{-0.5em}
  \caption{Runtime of \sn and baselines on limit queries with predicates. \sn's
  outperforms all baselines, even when including the training time of the proxy
  model. \sn especially outperforms baselines when the selectivity is high:
  random sampling will perform especially poorly on rare events.}
  \vspace{-0.5em}
  \label{fig:scrubbing-color}
\end{figure}

\begin{sloppypar}
As shown in Figure~\ref{fig:scrubbing-color}, \sn outperforms all baselines by
up to 300$\times$, even when including the proxy model training time. \sn
especially outperforms baselines on queries that have few matches, as random
sampling and \noscope will perform poorly in these settings.
\end{sloppypar}

%% file: tex/eval_spec.tex
\subsection{Specialized Neural Networks}

\begin{figure}[t]
  \centering
  \includegraphics[width=0.99\columnwidth]{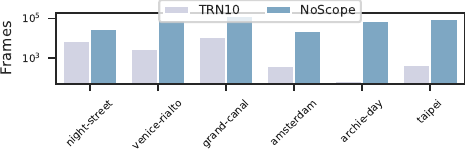}
  \vspace{-0.5em}
  \caption{Number of samples to find 10 of the requested objects for each query,
  using TRN10 or a representative \noscope NN. As shown, TRN10 significantly
  outperforms the \noscope NN on all videos. The y-axis is on a log-scale.
  Average of 3 runs.}
  \label{fig:tiny-scrubbing}
  \vspace{-0.5em}
\end{figure}

\begin{figure}[t]
  \centering
  \includegraphics[width=0.99\columnwidth]{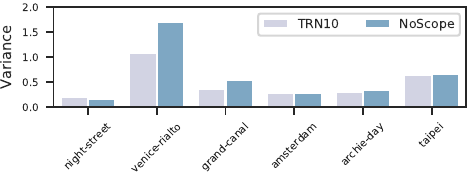}
  \vspace{-0.5em}
  \caption{Variance of the control variates estimator when using TRN10 or a
  representative \noscope NN (lower is better). As shown, TRN10 typically
  matches or beats the \noscope NN, except for \texttt{night-street}.}
  \label{fig:tiny-agg}
  \vspace{-0.5em}
\end{figure}

\minihead{Effect of NN type}
In this work, we used a tiny ResNet (referred to as TRN10) as the default
specialized architecture. ResNets are an extremely popular
architecture~\cite{coleman2017dawnbench, coleman2018analysis}. To test our
hypothesis that TRN10 is a good default, we compared TRN10 to a representative
\noscope NN~\cite{kang2017noscope}, parameterized by 32 base filters, 32 dense
neurons, and 4 layers.

We used TRN10 and the \noscope NN on limit queries for each of the
videos and computed the number of samples required to find the requested number
of events in Table~\ref{table:scrubbing-instances}. As shown in
Figure~\ref{fig:tiny-scrubbing}, TRN10 requires significantly fewer samples
compared to the \noscope NN on all videos.

We additionally used TRN10 and the \noscope NN for the aggregation tasks for
each video and computed the variance of the control variate estimator (the
variance of the estimator is directly related to the number of samples; lower is
better).  As shown in Figure~\ref{fig:tiny-agg}, TRN10 typically matches or
beats the \noscope NN, except for \texttt{night-street}.


\begin{figure}[t]
  \centering
  \includegraphics[width=0.99\columnwidth]{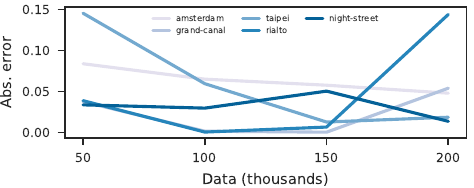}
  \vspace{-0.5em}
  \caption{Effect of the amount of data on the error of specialized NNs on the
  aggregation queries used in Section~\ref{sec:eval_aggregate}. As shown, the
  error tends to decrease until 150k training examples.}
  \label{fig:spec-data}
  \vspace{-0.5em}
\end{figure}

\minihead{Effect of training data}
To see the effect of the amount of training data on aggregate query performance,
we plotted the error of the specialized NNs on the aggregation queries used in
Section~\ref{sec:eval_aggregate}. We show results in Figure~\ref{fig:spec-data}.
As shown, the error tends to decrease until around 150,000 training examples and
levels off or increases, potentially due to overfitting.

\begin{figure}[t]
  \centering
  \includegraphics[width=0.99\columnwidth]{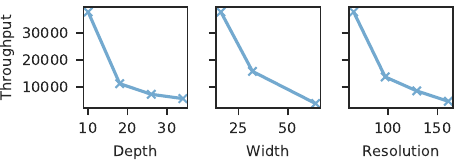}
  \vspace{-0.5em}
  \caption{Effect of the width, depth, and input resolution on the throughput of
  the tiny ResNet architecture. The throughput is proportional to the inverse of
  the width and the depth, and generally proportional to the inverse of the
  square of the input resolution.}
  \label{fig:spec-perf}
  \vspace{-0.5em}
\end{figure}

\minihead{Performance benchmarks}
We plotted the performance of tiny ResNets as the depth, width, and resolution
of the network and inputs varied. As shown in Figure~\ref{fig:spec-perf}, the
throughput of the tiny ResNets decreases linearly with depth and width. The
throughput generally decreases with the square of the resolution, but reduces
further if the maximum batch size that the GPU can fit decreases.

%% file: tex/related_work.tex
\section{Related Work}
\label{sec:related_work}

\sn builds on research in data management for multimedia and video, and
on recent advances in computer vision. We outline some relevant parts of the
literature below.

\minihead{AQP}
In AQP systems, the result of a query is returned significantly faster by
subsampling the data~\cite{garofalakis2001approximate}. Typically, the user
specifies an error bound~\cite{agarwal2013blinkdb}, or the error bound is
refined over time~\cite{hellerstein1997online}. Prior work has leveraged various
sampling methods~\cite{chaudhuri2007optimized, agarwal2014knowing},
histograms~\cite{acharya1999aqua, piatetsky1984accurate, greenwald2001space,
cormode2006space}, and sketches~\cite{henzinger1998computing,
jin2003dynamically, charikar2002finding}.

The key different in \sn is difference in cost of tuple materialization:
materializing a tuple for video analytics (i.e., executing object detection) is
orders of magnitude more expensive than in standard databases. To address this
challenge, we introduce a new form of variance reduction in the form of control
variates~\cite{hammersley1964general} via specialized NNs. This form of variance
reduction, and others involving auxiliary variables, does not apply in a
traditional relational database due to the cost imbalance.

\minihead{Visual data management}
Visual data management has aimed to organize and query visual data, starting
from systems such as Chabot~\cite{ogle1995chabot} and
QBIC~\cite{flickner1995query}. These systems were followed by a range of
``multimedia" database for storing~\cite{arman1993image,lee2005strg},
querying~\cite{aref2003video,la1996jacob,oh2000efficient}, and
managing~\cite{gibbs1993audio,jain1994metadata,yoshitaka1999survey} video data.
The literature also contains many proposals for query languages for visual
data~\cite{hwang1996querying, demir2011flexible, lu2015svql}; we discuss
how \fql differs from these languages in an extended version of this
paper~\cite{kang2019blazeit}.

Many of these systems and languages use classic computer vision techniques such as
\emph{low-level} image features (e.g., color) and rely on textual
annotations for semantic queries. However, recent advances in computer vision
allow the \emph{automatic} population of semantic data and thus we believe it is
critical to reinvestigate these systems. In this work, we explicitly choose to
extend SQL in \fql and focus on how these fields can be
\emph{automatically populated} rather than the syntax.

\minihead{Modern video analytics}
Systems builders have created video analytics systems, e.g.,
\noscope~\cite{kang2017noscope}, a highly tuned pipeline for binary detection:
it returns the presence or absence of a particular object class in video. Other
systems, e.g., \textsc{Focus}~\cite{hsieh2018focus} and
\textsc{Tahoma}~\cite{anderson2018predicate}, also optimize binary
detection. However, these systems are inflexible and cannot adapt to user's
queries. Additionally, as \noscope does not focus on the exploratory setting, it
does not optimize the training time of specialized NNs. In \sn, we extend
specialization and present novel optimizations for aggregation and limit
queries, which these systems do not support.

Other contemporary work use filters with a false negative rate (called
probabilistic predicates) that are automatically learned from a hold-out
set~\cite{lu2018accelerating}. These could be incorporated into \sn for
selection queries.

\begin{sloppypar}
Other systems aim to reduce latency of live queries (e.g.,
VideoStorm~\cite{zhang2017live}) or increase the throughput of batch analytics
queries (e.g., \textsc{Scanner}~\cite{poms2018scanner}) that are pre-defined
\emph{as a computation graph}. As the computation is specified as a black-box,
these systems do not have access to the semantics of the computation to perform
certain optimizations, such as in \sn. In \sn, we introduce \fql and an
optimizer that can infer optimizations from the given query. Additionally, \sn
could be integrated with VideoStorm for live analytics or \textsc{Scanner} for
scale-out.
\end{sloppypar}

We presented a preliminary version of \sn as a non-archival
demonstration~\cite{kang2019challenges}.

\minihead{Speeding up deep networks}
We briefly discuss two of the many forms of improving deep network efficiency.

First, a large body of work changes the NN architecture or weights for improved
inference efficiency, that preserve the full generality of these NNs. Model
compression uses a variety of techniques from pruning~\cite{han2015deep} to
compressing~\cite{chen2015compressing} weights from the original NN, which
can be amenable to hardware acceleration~\cite{han2016eie}. Model distillation
uses a large NN to train a smaller NN~\cite{hinton2015distilling}. These
methods are largely orthogonal to \sn, and reducing the cost of object detection
would also improve \sn's runtime.

\begin{sloppypar}
Second, specialization~\cite{kang2017noscope,shen2016fast} aims to improve
inference speeds by training a small NN to mimic a larger NN \emph{on
a reduced task}. Specialization has typically been applied in
\emph{specific} pipelines, e.g., for binary detection. In \sn, we
extend specialization to counting and multi-class classification. Further, we
show to how use specialized NNs as control variates and for limit queries.
\end{sloppypar}

%% file: tex/conclusion.tex
\section{Conclusions}

\begin{sloppypar}
Querying video for semantic information has become possible with advances in
computer vision. However, these NNs run up to 10$\times$ slower than real-time
and requires complex programming with low-level libraries to deploy.  In
response, we present \sn, a optimizing video analytics system with a declarative
language, \fql.  We introduce two novel optimizations for aggregation and limit
queries, which are not supported by prior work. These techniques can run orders
of magnitude faster than baselines while retaining accuracy guarantees, despite
potentially inaccurate specialized NNs. These results suggest that new classes
of queries can be answered over large video datasets with orders of magnitude
lower computational cost.
\end{sloppypar}

%% file: appendix.tex
\begin{appendix}

\section{\fql}
\label{appendix:fql}

\subsection{Populating \fql}
\label{appendix:populating-fql}

In this work, we focus on the batch analytics setting (as opposed to streaming
analytics) so we assume the all the video is available for processing. To our
knowledge, the majority of object detection and entity resolution methods are
deterministic. Thus, once the object detection and entity resolution methods are
fixed, the \fql tuples do not change.

\begin{figure}[t!]
  \centering
  \includegraphics[width=0.99\columnwidth]{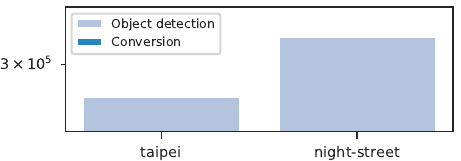}
  \vspace{-0.5em}
  \caption{
  Time of object detection and conversion to \fql tuples, when processing the
  entire video. Zoomed in and log-scaled for visibility.
  }
  \vspace{-0.5em}
  \label{fig:conversion}
\end{figure}

We have empirically found that, even for busy scenes, a fully populated \fql
table has around 100k rows per hour of video. We show the breakdown of time for
detection and processing the data to \fql tuples for \texttt{taipei} and
\texttt{night-street} in Figure~\ref{fig:conversion} (\texttt{taipei} is the
busiest video and thus takes the longest time to convert the output of object
detection to \fql tuples). As converting the output of object detection to \fql
tuples is a tiny fraction of total time, we ignore the transformation time.

\subsection{Other Visual Query Languages}

There are many other visual query languages in the literature, including
SVQL~\cite{lu2015svql}, CVQL~\cite{kuo2000content},
BVQL~\cite{donderler2005bilvideo}, and SRQL~\cite{le2008query}. \fql shares
similarities with many of these query languages, but has several key
differences. First, the majority of prior visual query languages either used
low-level features, assumed the semantic data was provided, or used classical
computer vision methods to extract semantic data. Second, many of these query
languages focuses on events or specific instances of objects (e.g., specific
actors in a movie). Third, many prior query languages implicitly focus on
relatively short videos (e.g., minutes) as opposed to the long-running video
streams we consider in this work.

In designing \fql, we specifically aimed to avoid the user having to annotate
video. Thus, we restricted the language to only allow expressions that can be
\emph{automatically} computed with high accuracy (e.g., we do not allow querying
for specific events). Researchers have only recently developed techniques to
automatically extract this information from video with high
accuracy~\cite{he2017mask}.

\subsection{\fql examples}
\label{appendix:fql-examples}

\begin{figure}[t!]
\centering
\begin{subfigure}[t]{0.45\columnwidth}
  \centering
  \small
\begin{verbatim}
SELECT COUNT
  (DISTINCT trackid)
FROM taipei
WHERE class = 'car'

\end{verbatim}
  \caption{Count distinct cars.}
  \label{query:distinct-cars}
\end{subfigure}\hspace{0.02\columnwidth}
\begin{subfigure}[t]{0.49\columnwidth}
  \centering
  \small
\begin{verbatim}
SELECT COUNT(*)
FROM taipei
WHERE class = 'car'
ERROR WITHIN 0.1
  CONFIDENCE 95%
\end{verbatim}
  \caption{Example of error rate.}
  \label{query:error}
\end{subfigure}

\vspace{0.4em}
\begin{subfigure}[t]{0.45\columnwidth}
  \centering
  \small
\begin{verbatim}
SELECT timestamp
FROM taipei
WHERE class = 'car'
FNR WITHIN 0.01
FPR WITHIN 0.01
\end{verbatim}
  \caption{Replicating \noscope.}
  \label{query:noscope}
\end{subfigure}\hspace{0.02\columnwidth}
\begin{subfigure}[t]{0.49\columnwidth}
  \centering
  \small
\begin{verbatim}
SELECT *
FROM taipei
WHERE class = 'car'
  AND classify(content)
    = 'sedan'
\end{verbatim}
  \caption{UDF to classify cars.}
  \label{query:subcars}
\end{subfigure}
\vspace{-0.5em}
\caption{Further examples of \fql.}
\vspace{-0.5em}
\label{fig:fql-ex-appendix}
\end{figure}

We give further examples of \fql in Figure~\ref{query:distinct-cars}.

First, counting the number of distinct cars can be written as in
Figure~\ref{query:distinct-cars}, which is not the same as counting the
average number of cars in a frame, as this query looks for distinct instances of
cars using \texttt{trackid} (cf.  Figure~\ref{query:aggregate}).
Second, error rates can be set using syntax similar to
BlinkDB~\cite{agarwal2013blinkdb}, as in Figure~\ref{query:error}.
Third, \noscope can be replicated in a similar manner to
Figure~\ref{query:noscope}.
Finally, a UDF could be used to classify cars into subtypes, as in
Figure~\ref{query:subcars}.

\subsection{Extensions to \fql}
\label{appendix:fql-extensions}

In this work, we focus on queries that can be optimized using current
techniques. However, many use cases require two features that we do not
currently support: 1) joins and 2) global identifiers for objects, which can
easily be added to the language. Thus, we describe how to add joins and global
identifiers to \fql.

\minihead{Joins}
Joins can be added to \fql as in standard SQL. Joins can be used to answer a
wider range of queries. For example, computing the average count of cars on a
single street camera can be used to determine when the given street is busiest,
but not when traffic is the highest across a city. This query could be answered
by joining on the timestamp and computing the average number of cars.

\minihead{Global identification}
While \fql can express a wide range of queries, several types of queries require
a unique global identifier. For example, if two cameras were placed a mile apart
on a highway, the flow rate of cars on the highway could be computed by the
average time it takes for a car to leave one camera and arrive at the other.

Thus, \fql could be extended by adding a field \texttt{globalid}. In the case of
cars, the license plate number could be used as a global identifier, but
computing \texttt{globalid} is a difficult task in general.

\section{Aggregation}
\label{appendix:aggregation}

\subsection{Full EBS algorithm}

\begin{algorithm}[t]
  $\mathrm{LB} \gets 0$

  $\mathrm{UB} \gets \infty$

  $t \gets 1$

  $k \gets 0$

  \While{LB + $\epsilon$ < UB - $\epsilon$}{
    $t \gets t + 1$

    \If{$t > \mathrm{floor}(\beta^k)$}{
      $k \gets k + 1$

      $\alpha \gets \mathrm{floor}(\beta^k) / \mathrm{floor}(\beta^{k-1})$

      $x \gets -\alpha \log d_k / 3$
    }

    $c_t \gets \bar{\sigma}_t \sqrt{2x/t} + 3 R x/t$

    $\mathrm{LB} \gets \max(LB, |\bar{X_t}| - c_t)$

    $\mathrm{UB} \gets \min(UB, |\bar{X_t}| + c_t)$
  }

  \caption{Full EBGStop algorithm that uses geometric sampling.}
  \label{alg:ebgstop}
\end{algorithm}

We first describe a simplified EBS algorithm for clarity and present the full
algorithm below. EBS will maintain a lower and upper bound of the
estimated quantity that always respects the confidence interval. Once the bounds
are within the error tolerance, the sampling stops. EBS requires an estimate of
the range of the data, which we estimate from the TMAS.

Formally, denote the error and confidence as $\epsilon$ and $\delta$. Denote the
sample mean and standard deviation as $\bar{X_t}$ and $\bar{\sigma}_t$ for the
$t$ samples respectively and $R$ to be the range of $X_t$. Let $d_t$ be a
sequence such that $\sum_{t=1}^{\infty} d_t \leq \delta$ and
\[
c_t = \bar{\sigma}_t \sqrt{\frac{2 \log(3/d_t)}{t}} + \frac{3 R \log(3/d_t)}{t}.
\]
Then, the upper and lower bounds are respectively
$\mathrm{UB} = \min_{1 \leq s \leq t | \bar{X}_s| - c_s}$.
and
$\mathrm{LB} = \max(0, \max_{1 \leq s \leq t} | \bar{X}_s| - c_s)$.
EBS will terminate when $\mathrm{LB} + \epsilon > \mathrm{UB} - \epsilon$.

\sn uses an improved EBS algorithm denoted EBGStop that uses geometric sampling.
We defer the proofs of correctness to~\cite{mnih2008empirical}. The full
algorithm is given in Algorithm~\ref{alg:ebgstop}. $\beta$ is a constant (we use
1.1) and we use $d_t = c/(\log_{\beta}(t))^p$ for $p=1.1$. Recall that
$\bar{X}_t$ and $\bar{\sigma}_t$ are the sample mean and standard deviation
respectively.

\subsection{Extensions to Aggregation}

In this work, we focus on extending aggregation beyond existing techniques, but
we describe how aggregation can optimize queries over other statistics.

\minihead{Aggregation for occupancy}
Another statistic of interest is the percentage of frames that are occupied,
which can be written in \fql as:
\begin{verbatim}
SELECT FCOUNT DISTINCT(*)
FROM taipei
WHERE class = 'car'
ERROR WITHIN 0.01 CONFIDENCE 95%
\end{verbatim}

To answer this query, \sn can perform Algorithm~\ref{alg:agg}, but instead train
a specialized NN that performs binary detection instead of counting the number
of objects in the frame.

\minihead{Aggregation for number of unique objects}
Another statistic of interest is the number of unique objects, which can be
written in \fql as in Figure~\ref{query:distinct-cars}.

To answer this query, \sn can use standard AQP with the following sampling
procedure:
\begin{itemize}
\item Sample $i$ iid from the number of frames.
\item For each sampled frame $f_i$, perform object detection and entity
resolution on frame $f_i$ and $f_{i+1}$.
\item Return the difference of the number of objects that are in frame $f_i$ but
not in $f_{i+i}$.
\end{itemize}

\minihead{Aggregation for box statistics}
Another class of statistics are statistics over the bounding boxes, e.g., an
analyst might be interested in the average box area or the average position. For
any numerical statistic over the bounding box, \sn can leverage traditional AQP,
using the following procedure:
\begin{itemize}
\item Sample a frame $f_i$ iid.
\item Perform object detection on $f_i$.
\item Sample iid a single box matching the predicates of the query.
\item Run the UDF over the box and return the answer.
\end{itemize}

To increase efficiency, \sn can cache the box information if the frame $f_i$ is
sampled again.

\section{Optimizing Content-based Selection}
\label{sec:filters}

\minihead{Overview}
In content-based selection, the user is interested information about the mask
or content of every instance of an event, e.g., finding red buses
(Figure~\ref{query:red_bus}). In these queries, the object detection method must
be called to obtain the mask. As object detection is the overwhelming
computational bottleneck, \sn aims to perform object detection as few times as
possible. For example, \sn can filter frames that lack red \emph{before}
performing object detection to look for buses.

To reduce the number of object detection invocations, \sn infers filters to
discard frames irrelevant to the query before running object detection on them.
\sn currently supports four classes of filters: 1) label-based filtering, 2)
content-based filtering, 3) temporal filtering, and 4) spatial filtering
(described in detail below).  Importantly, these filter types and parameters are
automatically selected from the query and training data as described below.

While some filters can be applied with no false negatives, others filters are
statistical in nature and may have some error rate. The error rate of these
filters can be estimated on a held-out set, as in
cross-validation~\cite{geisser2017predictive}. However, as prior
work~\cite{kang2017noscope} has considered how to set these error rates, we
only consider the case where the filters are set to have no false negatives on
the held-out set. Assuming the held-out set is representative of the unseen data
(i.e., no model drift, see \S\ref{sec:limitations}), this procedure will incur
few false negatives on the unseen data, which is sufficient for exploratory
queries or \texttt{LIMIT} queries.

\minihead{Physical Operators}
We present instantiations of each class of filter to demonstrate their
effectiveness. We describe each class of filter and \sn's instantiations of the
filter class.

\miniheadit{Label-based filtering}
In label-based filtering, the video is filtered based on the desired labels.
We leverage similar techniques to \noscope~\cite{kang2017noscope}, namely
training a specialized NN to discard frames without the label.

\miniheadit{Content-based filtering}
In content-based filtering, the video is filtered based on UDFs that return
continuous values (e.g., the UDF \texttt{redness} returns a measure of redness
of an image). \sn will attempt to learn a threshold based on the training data
or fall back to naive application of the UDF if it cannot learn a beneficial
threshold.

Currently, \sn supports automatically learning a filter from any UDF that only
takes the box content, by applying the UDF over the entire frame. For example,
if the query filters for \texttt{redness > R} and \texttt{area > A}, then
\texttt{redness(frame) > A $\cdot$ R} is a safe threshold. In practice, even
more conservative thresholds may work depending on the UDFs and video.

We describe when content-based filtering is effective below.

\miniheadit{Temporal filtering}
In temporal filtering, the video is filtered based on temporal cues. For
example, the analyst may want to find buses in the scene for at least $K$
frames. In this case, \sn subsamples the video at a rate of $\frac{K-1}{2}$. \sn
additionally support basic forms of filtering such as range-based queries. 

\miniheadit{Spatial filtering}
In spatial filtering, only regions of interest (ROIs) of the scene are
considered. For example, a street may have cars parked on the side but the
analyst may only be interested in vehicles in transit, so the analyst specifies
in the query which parts of the scene contain moving vehicles. The ROI is
specified by the user and can be used for faster object detection inference,
and activity outside the ROI can be ignored, which can increase the selectivity
of other filters.

\sn crops the images to be more square if the given ROI allows such an
operation. For example, if the query only looks for objects with
\texttt{xmax(mask) < 720} in a 1280$\times$720 video, \sn will resize the frames
to be 720$\times$720. Standard object detectors run faster when the input is
more square: in most existing detectors, the input image is resized so that the
short-edge is a specific size and the aspect ratio is held
constant~\cite{ren2015faster,he2017mask} (for a fixed short-edge size, reducing
the long-edge size will make the image smaller).  As the computation scales with
the resolution, square images result in the least computation.

\minihead{Operator Selection}
\sn will infer which filters can be applied from the user's query. We describe
how each class of filter can be inferred from the query.

First, if the user selects an area of the video, \sn resizes the frame to be as
square as possible, while keeping the area (along with some padding) visible, as
described above.

Second, \sn infers the times in the video and the subsampling rate from the
query to achieve exact results. For example, if the user queries for objects
present in the video for at least 30 frames (1 second), \sn can sample once very
14 frames.

Third, if the user selects classes, \sn trains a specialized
NN to detect these classes, as in \noscope. Then, \sn estimates the threshold on
unseen data to attempt no false negatives.

Fourth, if the user provides a content-based UDF over the content (e.g.,
to determine the color of the object), \sn can apply the UDF over the entire
frame (as opposed to the box), and filter frames that do not satisfy the UDF at
the frame level. \sn sets UDF filter thresholds similar to how it sets
thresholds for specialized NNs. However, for this procedure to be effective, the
UDF must return a continuous value that can be scaled to a confidence.  Consider
two possible UDFs for redness: 1) a UDF which returns true if the over 80\% of
the pixels have a red-channel value of at least 200 (out of 256) and 2) a UDF
that returns the average of the red-channel values. While both UDFs can be used
as a filter, the second will be more effective.

\minihead{Correctness}
\sn will perform object detection on all frames that pass its filters, so no
false positives will be returned. Spatial and temporal filters are exact, but
label-based and content-based filters are not. For probabilistic filters, we
set the thresholds so there are no false negatives on the held-out set. The
accuracy will be high, but possibly not perfect, and will return no
false positives for \texttt{LIMIT} queries.

\minihead{Time complexity}
Before filter $i$ is applied, denote the remaining frames $F_i$. Denote the cost
of filter $i$ on a frame to be $c_i$ and the cost of object detection to be
$c_o$. If $c_i \cdot F_i < c_o \cdot (F_i - F_{i+1})$, then the filter is
beneficial to run. The filters we consider in this work are 3 to 6 orders of
magnitude faster than object detection and are thus nearly always worth running.

\section{Further Experiments and Experimental Details}
\label{appendix:experiments}

\subsection{Further Experimental Details}

\begin{figure}[t!]
  \centering
  \begin{verbatim}
SELECT *
FROM VIDEO
WHERE class = OBJ
  AND area(mask) > AREA
  AND COLOR(content) > COLOR_VAL
GROUP BY timestamp
HAVING SUM(class = OBJ) >= NBOBJ
LIMIT NFRAMES
GAP 300
  \end{verbatim}
  \vspace{-1.5em}
  \caption{Base query for limit queries with predicates. The exact values for
  the variables are shown in Table~\ref{table:limit-predicate}.}
  \label{fig:limit-predicate}
\end{figure}

\begin{table*}[t!]
  \centering
  \caption{Query details for limit queries with predicates tested in this work.}
  \label{table:limit-predicate}
  \setlength\itemsep{2em}
  \begin{tabular}{cccccccc}
  Video name & Object & Number of & Number of & Area & Color type & Color value & Instances \\
             &        & objects   & frames \\
  \hline
  \texttt{taipei} & bus & 2 & 5 & 55000 & Background red & 13.0 & 5 \\
  \texttt{night-street} & car & 2 & 10 & 90000 & Luma & 110.0 & 14 \\
  \texttt{rialto} & boat & 3 & 5 & 40000 & Luma & 110.0 & 8 \\
  \texttt{grand-canal} & boat & 2 & 5 & 60000 & Background blue & 6.5 & 6 \\
  \texttt{amsterdam} & car & 2 & 5 & 18000 & Background blue & 15.0 & 5 \\
  \texttt{archie} & car & 2 & 10 & 100000 & Absolute red & 15.0 & 20
  \end{tabular}
\end{table*}

\minihead{Query details for limit queries with predicates}
We provide query details for the limit queries with predicates. The base query
is shown in Figure~\ref{fig:limit-predicate}. The exact values for the variables
are shown in Table~\ref{table:limit-predicate}. As shown, each query searched
for 5 or 10 instances of $N$ objects with at least a certain area and color
level.

\begin{figure}[t!]
  \centering
  \begin{verbatim}
SELECT FCOUNT(*)
FROM VIDEO
WHERE class = OBJ
  AND area(mask) > AREA
  AND COLOR(content) > COLOR_VAL
ERROR WITHIN 0.001
AT CONFIDENCE 95%
  \end{verbatim}
  \vspace{-1.5em}
  \caption{Base query for aggregation queries with predicates. The exact values for
  the variables are shown in Table~\ref{table:limit-predicate}.}
  \label{fig:agg-predicate}
\end{figure}

\minihead{Query details for aggregation queries with predicates}
We show the base query for aggregates with predicates in
Figure~\ref{fig:agg-predicate}. The exact values for the variables are the same
as the limit queries, with details shown in Table~\ref{table:limit-predicate}.

\begin{table}[t!]
  \centering
  \caption{Number of samples required for the aggregation queries in
  Figure~\ref{fig:agg-bypass} using naive AQP. The target error was 0.1.}
  \label{table:aqp-nb-samples}
  \setlength\itemsep{2em}
  \begin{tabular}{lc}
  Video Name            & Number of samples \\ \hline
  \texttt{taipei}       &  3088 \\
  \texttt{night-street} &  1971 \\
  \texttt{rialto}       &  3826 \\
  \texttt{grand-canal}  &  2107 \\
  \texttt{amsterdam}    &  2018
  \end{tabular}
\end{table}

\minihead{Number of samples for aggregation queries}
We report the number of samples required for the naive AQP baseline in the
aggregation queries shown in Figure~\ref{fig:agg-bypass}. The target error rate
was 0.1. The number of samples are shown in Table~\ref{table:aqp-nb-samples}.

\begin{table}[t!]
  \centering
  \caption{Number of times the confidence interval was respected for the
  aggregation queries in Figure~\ref{fig:agg-bypass} using EBS, out of a total
  of 100 runs. The target
  error was 0.01, the most stringent we consider.}
  \label{table:aqp-confidence}
  \setlength\itemsep{2em}
  \begin{tabular}{lc}
  Video Name            & Within error tolerance \\ \hline
  \texttt{taipei}       &  99 \\
  \texttt{night-street} & 100 \\
  \texttt{rialto}       & 100 \\
  \texttt{grand-canal}  & 100 \\
  \texttt{amsterdam}    & 100 \\
  \texttt{archie-day}   & 100
  \end{tabular}
\end{table}

\subsection{Empirical Analysis of EBS}
We empirically verify that estimating the range of the data does not affect the
confidence intervals when using EBS. We run the EBS sampling procedure 100 times
for each video and show the number of times the confidence interval ($\sigma =
95\%$) was respected in Table~\ref{table:aqp-confidence} at the most stringent
error tolerance we consider ($\epsilon = 0.01$). As shown, every query respects
the confidence interval.

\subsection{Experiments with Full Frames}

In the experiments in the main text, we filtered portions of frames where object
detection performed poorly. To verify that this choice does not affect \sn's
performance, we perform additional experiments for cardinality-limited queries
and sampling queries. The queries are similar to the ones performed in
Figures~\ref{fig:agg-control} and \ref{fig:scrubbing-all}.

\begin{figure}[t!]
  \centering
  \includegraphics[width=0.99\columnwidth]{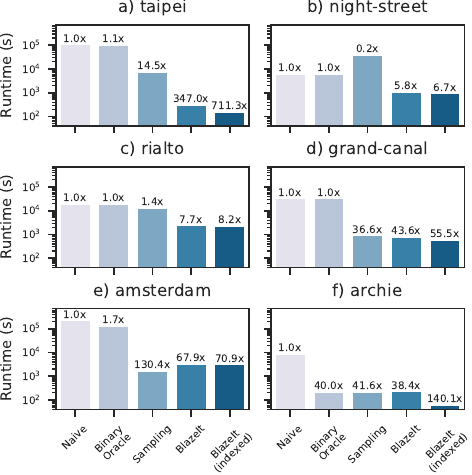}
  \caption{End-to-end runtime of baselines and \sn on limit queries without
  filtering portions of frames. The y-axis is log-scaled. All queries looked for
  10 events. As shown, \sn outperforms the baselines except for sampling on
  \texttt{amsterdam}.}
  \label{fig:scrubbing-conf}
\end{figure}

\begin{table}[t!]
  \centering
  \caption{Percent reduction in the number of samples necessary to answer an
  aggregation query with an error tolerance of 0.01. These experiments were
  conducted without filtering portions of frames. As shown, control variates
  still provide a reduction in all cases, but the relative speedups are data
  dependent.}
  \label{table:cv-conf}
  \setlength\itemsep{2em}
  \begin{tabular}{lc}
  Video Name            & \% reduction \\ \hline
  \texttt{taipei}       & 19.0\% \\
  \texttt{night-street} & 22.5\% \\
  \texttt{rialto}       & 3.4\% \\
  \texttt{grand-canal}  & 0.4\% \\
  \texttt{amsterdam}    & 6.1\% \\
  \texttt{archie-day}   & 3.7\%
  \end{tabular}
\end{table}

We show results in Figure~\ref{fig:scrubbing-conf} and
Table~\ref{table:cv-conf}. As shown, \sn still outperforms all baselines except
for on \texttt{amsterdam} for the cardinality-limited selection query.

\input{eval_spatiotemporal.tex}

%
%
%
%


\section{Rule-based Optimizer}
\sn currently uses a rule-based optimizer for queries, as we have found it
sufficient to optimize a large class of queries. \sn's rule-based optimizer
attempts to classify queries as (approximate) aggregation, cardinality-limited
queries, or content-based selection and applies the rules described in the
main paper in these settings. For all other queries, \sn falls back to
materializing all the rows in the \fql table.

To determine if a query is an approximate aggregation query (exact aggregation
requires all the relevant rows to be materialized), \sn inspects the \fql query
for the \texttt{ERROR} keyword and an aggregation keyword (e.g.,
\texttt{FCOUNT} or \texttt{AVG}). \sn will then determine the necessary fields
and perform approximate aggregation as described above.

To determine if a query is a cardinality-limited query, \sn inspects the \fql
query for the \texttt{LIMIT} keyword. \sn then determines the necessary fields
for the query and executes the cardinality-limited query as described above.

To determine if the query is a content-based selection query (with applicable
filters), \sn will inspect the query for the predicates as described in
Section~\ref{sec:filters} and apply them as described above.

In all other cases, \sn will default to applying object detection to every
frame.

\end{appendix}

%% file: eval_spatiotemporal.tex
\subsection{Content-based Selection Queries}
\label{sec:eval_selection}
To illustrate the effectiveness of content-based filters, we evaluate \sn on the
query shown in Figure~\ref{query:red_bus}.

We run the following variants:
\begin{itemize}
  \setlength\itemsep{-0.0em}
  \item Naive: we perform object detection on every frame.
  \item Binary oracle: we perform object detection on the frames that
  contain the object class of interest.
  \item \sn: we apply the filters described in \S\ref{sec:filters}.
\end{itemize}
We do not include an AQP baseline, as sampling does not help for exhaustive queries.

For each query, \sn's CBO trains, estimates the selectivity, and computes the
threshold for each filter applicable to the query (which is determined by \sn's
rule-based optimizer). We include the time to train the filters and select the
thresholds in the runtime. Due to the large computational cost of running the
object detector, we extrapolate its cost by multiplying the number of calls by
the runtime of the object detector.

\begin{figure}[t!]
  \centering
  \includegraphics[width=0.99\columnwidth]{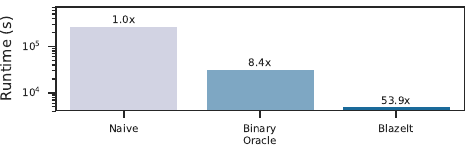}
  \vspace{-1em}
  \caption{
  End-to-end throughput of baselines and \sn on the query in
  Figure~\ref{query:red_bus}. The y-axis log scaled.
  }
  \vspace{-0.5em}
  \label{fig:spatiotemporal}
\end{figure}

\minihead{End-to-end performance}
The results for the end-to-end runtime of the baselines
and \sn are shown in Figure~\ref{fig:spatiotemporal}. As buses are
relatively rare (12\% occupancy, see Table~\ref{table:videos}), the binary
oracle
performs well on this query, giving a 8.4$\times$ performance improvement over
the naive method. However, \sn outperforms the binary oracle by 6.4$\times$,
due to its extended classes of filters. Furthermore, \sn delivers up to
54$\times$ improved throughput over naive methods for this query.

\begin{figure}[t!]
  \centering
  \includegraphics[width=0.99\columnwidth]{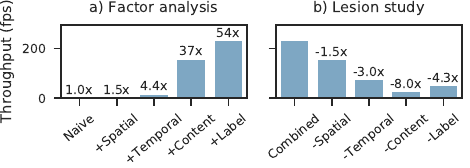}
  \vspace{-1em}
  \caption{
  Factor analysis and lesion study of \sn's filters on the query in
  Figure~\ref{query:red_bus}.
  }
  \vspace{-0.5em}
  \label{fig:st-factor}
\end{figure}

\minihead{Factor analysis}
We perform a factor analysis (adding filters one at a time) and lesion study
(individually removing filters) to understand the impact of each
class of filter. Results are shown in
Figure~\ref{fig:st-factor}. As shown in the factor analysis, every filter adds a
non-trivial speedup. Additionally, removing any class of filter reduces
performance. Thus, every class of filter improves performance for this query.